\documentclass[useAMS,usenatbib,letterpaper]{mn2e}
\usepackage[centering,totalwidth=480pt,totalheight=680pt]{geometry}
\usepackage{amsmath}
\usepackage{nccmath} 
\usepackage{stfloats}
\usepackage{epsfig}
\usepackage{natbib}
\usepackage{subfigure}
\usepackage{multirow}
\usepackage{amssymb}
\usepackage{gensymb} 
\usepackage{graphicx}

\def \hi 		{$\emph{h}^{-1}$}

\newcommand{\vvir}{V_{\rm vir}}
\newcommand{\rvir}{r_{\rm vir}}
\newcommand{\cool}{{\rm cool}}
\newcommand{\gas}{{\rm gas}}
\newcommand{\hot}{{\rm hot}}
\newcommand{\msun}{\rm M_{\odot}}
\newcommand{\Cint}{C_{\rm int}}
\newcommand{\Crad}{C_{\rm rad}}
\newcommand{\Csig}{C_{\sigma}}
\newcommand{\LCDM}{$\Lambda$CDM}

\DeclareMathOperator\erf{erf}
\def \swidth  {0.85\textwidth}

\title{Understanding the Structural Scaling Relations of Early-Type Galaxies}
\author[Porter et al.]{L.~A. Porter$^{1,2}$, R.~S. Somerville$^{3}\footnotemark[1]$,  J.~R. Primack$^{1,2}$, and P.~H. Johansson$^{4}$\\
 $^1$Department of Physics, University of California, Santa Cruz, California 95064, USA\\
	$^2$Santa Cruz Institute for Particle Physics, University of California, Santa Cruz, California 95064, USA\\
     $^3$Department of Physics and Astronomy, Rutgers University, Piscataway, New Jersey 08854, USA\\
     $^4$ Department of Physics, University of Helsinki, Gustaf H\"allstr\"omin katu 2a,
FI-00014 Helsinki, Finland\\
}

\begin{document}
\maketitle
\begin{abstract}

We use a large suite of hydrodynamical simulations of binary galaxy
mergers to construct and calibrate a physical prescription for computing
the effective radii and velocity dispersions of spheroids.  We
implement this prescription within a semi-analytic model embedded in
merger trees extracted from the Bolshoi $\Lambda$CDM N-body
simulation, accounting for spheroid growth via major and minor mergers
as well as disk instabilities.  We find that without disk
instabilities, our model does not predict sufficient numbers of
intermediate mass early-type galaxies in the local universe. Spheroids
also form earlier in models with spheroid growth via disk
instabilities.  Our model correctly predicts the normalization, slope,
and scatter of the low-redshift size-mass and Fundamental Plane
relations for early type galaxies. It predicts a degree of curvature
in the Faber-Jackson relation that is not seen in local observations,
but this could be alleviated if higher mass spheroids have more
bottom-heavy initial mass functions.  The model also correctly
predicts the observed strong evolution of the size-mass relation for
spheroids out to higher redshifts, as well as the slower evolution in
the normalization of the Faber-Jackson relation.  We emphasize that
these are genuine predictions of the model since it was tuned to match
hydrodynamical simulations and not these observations.
\end{abstract}
\begin{keywords}
galaxies: interactions -- galaxies: evolution -- galaxies: elliptical and lenticular, cD -- galaxies: formation
\end{keywords}
	\footnotetext[1]{email: somerville@physics.rutgers.edu}
\section{Introduction}

One of the most striking and well-known aspects of the galaxy
population in the local universe is the distinction between `early
type', or spheroid-dominated, galaxies, and `late type', or
disk-dominated galaxies.  Classical early-type galaxies (``giant
ellipticals'') in the local universe are dominated by random motions,
have compact, concentrated light profiles, and are typically red and
gas poor. Late-type galaxies are rotation supported, have more
extended light profiles, and tend to be gas-rich, blue, and star
forming. While local galaxy populations span a continuum in all of
these characteristics, some properties (most dramatically color or
specific star formation rate) show a pronounced bimodality
\citep[e.g.][and references therein]{baldry:04,blanton:09}.

In addition, both early- and late-type galaxies obey qualitatively
similar, yet distinct, \emph{scaling relations} between their
kinematic and structural properties, and mass or luminosity. For
example, elliptical galaxies obey a relation between surface
brightness, size, and velocity dispersion
\citep{Djorgovski:1987b,Dressler:1987a,Faber:1987a}, termed the
Fundamental Plane (FP).  This plane is tilted from the plane one would
expect from a simple application of the virial theorem, indicating
that further processes, such as non-homology or a varying
mass-to-light ratio, must play a role \citep{Jorgensen:1996b}.
Projections of this relationship form the familiar Faber-Jackson
relation \citep{Faber:1976a} between luminosity (or stellar mass) and
velocity dispersion, and the Kormendy relation between luminosity or
stellar mass and radius \citep{Shen:2003a}. Disk galaxies show similar
scaling relations, but the size-mass relation for disks has a
different slope, such that spheroid-dominated galaxies are smaller at
fixed mass than their disk-dominated counterparts
\citep{Shen:2003a}. Furthermore, there is a clear correlation between
galaxy kinematic and structural properties and star formation history:
more massive, higher velocity dispersion early-type galaxies have
older, more metal rich stellar populations \citep{Gallazzi:2006a}.
The bulges of spiral galaxies show similar scaling relations and
correlations, hinting that the formation mechanism of at least
``classical'' bulges may be the same as that of elliptical galaxies
\citep{burstein:97}.

Only recently have large, multi-wavelength surveys begun to
characterize the \emph{evolution} of galaxy demographics and scaling
relations over a significant span of cosmic history.  These surveys
have shown that the stellar mass contained in red sequence (passive)
galaxies has increased significantly since $z\sim 1$, while the mass
in blue (actively star forming) galaxies has stayed approximately
constant \citep{bell:07,faber:07}. This implies that active galaxies
have been \emph{transformed} into passive galaxies. Moreover, the gap
between the colors of active and passive galaxies in the local
universe (the ``green valley'') implies that this tranformation must
have been fairly rapid, giving rise to the frequently heard statement
that star formation is being \emph{quenched} in these systems. These
studies have recently been extended to higher redshifts ($z\sim
2$--3), showing a continuation of this trend
\citep{brammer:11,Muzzin:2013}.

In addition, extensive imaging with the Hubble Space Telescope has
enabled the study of galaxy structural properties and stellar
populations in the rest-frame optical out to $z\sim2$. An unexpected
population of extremely compact, yet massive, galaxies has been
discovered at $z\sim 2$--3, which is extremely rare in the local
Universe \citep{daddi:05,trujillo:07,vandokkum:08,Cassata:2013}, raising
the question of the nature of the descendants of these objects. The
structural scaling relations for disks and spheroids also
evolve over cosmic time, and the size-mass relation for early-types
evolves much more rapidly with redshift than that of their disk
dominated counterparts
\citep{Trujillo:2006a,Buitrago:2008a,Williams:2010b}.  This rapid
increase in size is accompanied by a smaller increase in central
velocity dispersion
\citep{Cappellari:2009a,Cenarro:2009a,Bezanson:2011a}, suggesting that
the cores of these galaxies are in place at high redshifts.
Recent observations have begun to probe the evolutionary link between
diffuse star-forming galaxies and compact quiescent galaxies at high
redshift.  These studies have found populations of dense galaxies with
high rates of star formation \citep{Wuyts:2011a,Barro:2013a}, as well
as compact post-starburst galaxies \citep{Whitaker:2012a} above $z
\sim 1$, suggesting that these galaxies are in the process of
undergoing morphological transformations and quenching.

Several different mechanisms have been proposed to explain this
transformation.  Hydrodynamical simulations have shown that major
mergers of gas-rich galaxies can induce massive amounts of star
formation, transforming a rotation-supported disk into a
pressure-supported bulge
\citep{Mihos:1994a,Mihos:1994b,Barnes:1996b,Naab:2006a,Cox:2008b}.
Early type galaxies can also be formed or enlarged through a sequence
of multiple gas-poor (dry) minor mergers \citep{Naab:2007a,Naab:2009a}.

A number of authors have proposed an evolutionary link between the
formation of a bulge, the growth of a central supermassive black hole
(SMBH), and quenching of massive galaxies
\citep[e.g.][]{Croton:2006a,Bower:2006a,Somerville:2008a,hopkins_cosmored:08}.
The same merger-induced torques that drive gas into galaxy centers,
fueling nuclear starbursts, may also drive accretion onto a SMBH,
leading to a quasar or Active Galactic Nucleus (AGN). This accretion
can drive powerful outflows that may clear out much of the cold gas
from the remnant, quenching further star formation
\citep{springel:05a,dimatteo:05}.  Once a massive SMBH is present,
even small amounts of accretion may be able to produce radio jets that
efficiently heat the surrounding hot gas halo, preventing future
cooling and star formation \citep[][and references
  therein]{fabian:12}.  It is interesting and suggestive, though far
from conclusive, that the galaxy merger rate estimated from
observations of close pairs and morphologically disturbed galaxies
appears to be consistent with what would be needed to account for the
growth of the quenched population, and the triggering of observed
bright AGN \citep{hopkins_cosmored:08,robaina:10,bundy:08}.

It has been suggested that bulges may also form and grow in-situ due
to internal gravitational instabilities. This process can take the
form of the formation of a bar that destabilizes the disk,
transferring mass into a compact, dynamically hot component
\citep{Toomre:1964a,hohl:71,OP:73,combes:90}. Clumps of gas may also
form in the disk, and migrate inwards
\citep{elmegreen:2008,Dekel:2009a,Bournaud:2011a} to form a bulge.
However, the efficiency of bulge formation via this in-situ channel,
the detailed physics of the process, and its importance relative to
mergers in a cosmological context, remain poorly understood.

A general picture has emerged in which the progenitors of early type
galaxies form at high redshift in one or more gas-rich ``wet''
mergers, and subsequently grow through predominantly dry mergers,
which can build up a more diffuse bulge, while leaving the center
relatively dense
\citep[e.g.,][]{Naab:2007a,Naab:2009a,hopkins_bundy:09,Oser:2010a,Johansson:2012b}. Such
a picture would be consistent with observations suggesting that
elliptical galaxies formed their centers rapidly, while the outer
regions were accreted over longer timescales
\citep{Dokkum:2010b,Forbes:2011a}. However, this picture has not been
tested quantitatively by confronting the predictions of large volume
cosmological simulations with observations of galaxy demographics and
scaling relations and their evolution over cosmic time.

In order to make detailed predictions regarding the connection between
star formation history and internal structure for galaxy populations,
it is necessary to simultaneously treat the cosmological framework of
the growth of structure through mergers and accretion, while
simultaneously resolving the internal structure and kinematics of
galaxies. One needs to know what kind of objects are merging, and
their gas content, morphology, structure, size, accretion rate, star
formation rate, etc. The dynamic range needed to simultaneously
resolve galaxy internal structure while simulating cosmological
volumes is currently difficult or impossible to achieve with purely
numerical hydrodynamic techniques.

Traditional semi-analytic models (SAMs) make predictions for the
global properties of galaxies as they form and evolve within the
framework of the \LCDM\ paradigm for structure formation.
\citep{Kauffmann:1993a,Cole:1994a,Somerville:1999b,Cole:2000a,Hatton:2003a,Croton:2006a,De-Lucia:2006a,Bower:2006a,Somerville:2008a,Fontanot:2009a,Benson:2010a,Cook:2010a,Guo:2011a,Somerville:2012a}.
SAMs have been shown to be a useful tool for examining the evolution
of galaxy morphological demographics and different channels for bulge
formation in a cosmological context. For example,
\citet{De-Lucia:2006a} and \citet{benson:10} have studied the build-up
of the spheroid-dominated population over cosmic time.
\citet{parry:09}, \citet{De-Lucia:2011a} and \citet{Fontanot:2012a}
have used SAMs to explore the relative importance of mergers and disk
instabilities in building the population of early-type galaxies.

However, most SAMs do not provide detailed predictions about the
internal structure of galaxies. A simple approach to predict the
radial sizes of disk-dominated galaxies, using angular momentum
conservation arguments \citep{Mo:1998a}, has been shown to be
surprisingly successful at reproducing the evolution of the size-mass
relationship for disk-dominated galaxies \citep{Somerville:2008b}. But
many previous attempts to model the radii and velocity dispersions of
early-type galaxies have been less successful.  \cite{Cole:2000a}
applied a simple formula to predict the radii of spheroidal remnants
following a major merger using the virial theorem and conservation of
energy, assuming that the merging galaxies have a similar structure to
the resulting spheroidal galaxy.  While this relation may be correct
for dissipationless gas-poor mergers, the energy lost due to star
formation in gas-rich mergers results in a deviation from the virial
relation, and smaller remnant radii
\citep{cox:06b,Dekel:2006a,Robertson:2006a}.

Incorporating this dissipation is probably essential: a study using
the \cite{Bower:2006a} SAM framework found that they were able to
match the observed $z\sim 0$ spheroid size-mass relation only by
including dissipation
\citep{Shankar:2010a,Shankar:2010b}. \citet{khochfar:06} have also
presented a semi-analytic model for the sizes of spheroids in which
the gas fraction of the progenitors played a key role in explaining
the observed size evolution and the scatter in the size-mass relation
at the present day. Another recent study based on SAMs that did not
include dissipation in modeling merger remnants produced a size-mass
relation with too shallow a slope and too large a scatter
\citep{Guo:2011a}.

A large body of literature has studied the structural properties of
binary merger remnants in N-body plus smoothed particle hydrodynamics
(SPH) simulations
\citep{cox:06b,Dekel:2006a,Robertson:2006a,Hopkins:2008a}. While these
studies have provided encouraging results regarding the explanation of
spheriod scaling relations, these simulations have a number of
significant limitations. They are not in a proper cosmological context
and do not include a hot gas halo or cosmological accretion of gas or
dark matter. The initial conditions are idealized and arbitrary, and
represent only single, binary mergers, predominantly major mergers of
fairly gas-rich disk-dominated progenitors. In contrast, cosmological
simulations predict that multiple mergers, mixed-morphology mergers,
gas-poor (``dry'') and minor mergers are all statistically important
in forming the observed present day population of early-type galaxies
\citep{moster:14,khochfar:03}.

Using a large suite of these binary SPH merger simulations
\citep{Cox:2004b,Cox:2006a,Cox:2008b}, Covington et al.~(2008,
hereafter C08) developed an analytic model to predict the effective
radius and velocity dispersion following the major merger of two disk
galaxies.  This model was based upon the virial theorem with the
incorporation of energy losses due to dissipation.  In further work,
Covington et al.~(2011, hereafter C11) simplified the model and
applied it via post-processing to mergers of disk-dominated galaxies
with properties taken from the \cite{Croton:2006a} Millennium SAM and
the \cite{Somerville:2008a} SAM.

C11 showed that the C08 model reproduces the steeper slope of the
size-mass relation of spheroid-dominated galaxies relative to disks,
as observed. In addition the model predicts a much smaller dispersion
in the size-mass relation for early-type galaxies relative to the
dispersion in the corresponding relationship for the progenitor disk
galaxies.  This occurs because the disk galaxies with larger radius
for their mass tend to be more diffuse and gas-rich, and a higher
progenitor gas content leads to more dissipation during the merger,
resulting in more compact remnants. In contrast, disk galaxies with
smaller radii tend to be gas poor, resulting in less dissipation and
less compact stellar spheroids. These effects conspire to produce
remnants that are of similar sizes, irrespective of the sizes of the
progenitor galaxies.  In addition, the C08 model qualitatively
reproduced the evolution of the size-mass relation with redshift
\citep{Trujillo:2006a}, and correctly reproduced a tilt in the FP away
from the simple virial relation.  Using the methods of C08 and C11, as
well as an alternative prescription in a similar spirit from
\cite{Hopkins:2009e}, \cite{Shankar:2013a} reached similar
conclusions.

In this paper, we develop a more complete model for predicting the
size and velocity dispersion of spheroids by augmenting the initial
suite of SPH merger simulations used by Covington and collaborators,
which were solely for fairly gas-rich, disk-dominated progenitors, to
include mergers involving spheroid-dominated and gas-poor progenitors
using the simulation suite presented in \cite{Johansson:2009a}. We
present an extended version of the C08 model that accounts for these
additional variables, and implement the new model self-consistently
within a full semi-analytic model. In addition, we implement a model
for the sizes and velocity dispersions of spheroids formed in disk
instabilities.
 
We make use of the ``Santa Cruz'' SAM, first presented in
\citet{Somerville:1999b} and significantly updated in
\citet[][S08]{Somerville:2008a} and \citet[][S12]{Somerville:2012a},
now run within merger trees extracted from the Bolshoi cosmological
N-body simulation \citep{Klypin:2011a,Trujillo-Gomez:2011a} using the
ROCKSTAR algorithm developed by \citet{Behroozi:2013a}.
The Santa Cruz SAM has been shown to be quite successful in predicting
many properties of nearby galaxies, including the stellar mass and
luminosity function \citep{Somerville:2008a,Somerville:2012a}, disk
gas fractions, the relative fraction of disk vs. spheroid-dominated
galaxies \citep{Hopkins:2009c}, and the fraction of active vs. passive
central galaxies as a function of stellar mass \citep{kimm:09}.  In
addition, the model reproduces the evolution of the size-mass
relationship for disk-dominated galaxies to $z\sim 2$
\citep{Somerville:2008b}, and is consistent with observational
constraints on the galaxy merger rate
\citep{Lotz:2011b}. \citet{Hirschmann:2012a} showed that a version of
the S08/S12 SAM with minor modifications also reproduced the evolution
of the luminosity function of radiatively efficient AGN from $z\sim
5$--0. \citet{Lu:2013} recently presented a comparison of three SAM
codes, including the same models used here, run within the same
Bolshoi-based merger trees, and showed broad agreement between the
predictions of the three models.
Therefore this SAM should provide a reasonably robust and
reliable framework within which to predict the evolution of the
properties of spheroids within a cosmological context.

The paper is arranged as follows.  In Section 2 we provide a brief
overview of all the physical processes in the SAM, paying particular
attention to the areas where the current SAM differs from
recently-published versions
\citep{Somerville:2008a,Somerville:2012a,Hirschmann:2012a}.  In
Section 3 we present our new analysis of hydrodynamical binary merger
simulations spanning a broader range of merger configurations and the
extended model that we use to predict the effective radii and velocity
dispersions of spheroids.  As there is still considerable uncertainty
in the importance of disk instabilities in producing spheroids, and
the details of how this process works, we present three models: one
without disk instabilities, one in which only the stellar disk
participates in the instability, and one in which both gas and stars
in the disk participate in the instability. In Section 4 we present
the stellar mass function, bulge-to-total ratio, size-mass,
Faber-Jackson, and Fundamental Plane relations for the simulated
galaxies, from redshift zero to $\sim 2$, and compare with available
observations.  We discuss the implications of our results and conclude
in Section 5.

\section{Methods}
Our baseline SAM is an extension of the model of
\citet{Somerville:1999b} and
\cite{Somerville:2001a,Somerville:2008a,Somerville:2012a}.  Galaxies
form, evolve, and merge in a hierarchical manner, following the growth
of their underlying dark matter halos.  We include prescriptions for
the radiative cooling of gas, star formation, supernova feedback,
black hole growth and AGN feedback, and chemical enrichment of the
stars, interstellar medium (ISM), and intergalactic medium (IGM).  We
provide a brief summary of the recipes used here, emphasizing any
relevant differences between this model and previous versions.  For
the full details of the SAM, we refer readers to
\citet{Somerville:2008a} and \citet{Somerville:2012a}.

In this work, we make use of merger trees extracted from the Bolshoi
N-Body dark matter simulation
\citep{Klypin:2011a,Trujillo-Gomez:2011a} using the ROCKSTAR method
developed by \citet{Behroozi:2013a}.  The simulation is complete down
to halos with virial velocity $V_{\rm circ} $ = 50 km/s, with a force
resolution of 1 \hi \ kpc and a mass resolution of $1.9 \times 10^{8}
\ \msun$ per particle.  Most previously published versions of the
Santa Cruz SAM were based on merger trees constructed using the
Extended Press-Schechter (EPS) method; merger trees extracted from
N-body simulations are presumably more accurate. However, we find that
running the SAM with the same parameters on the EPS and Bolshoi based
merger trees yields very similar results.

All results presented here assume a $\Lambda$CDM cosmology, with
cosmological parameters chosen to match those adopted in the Bolshoi
simulation: $\Omega_m$ = 0.27, $\Omega_\Lambda$ = 0.73, $\emph{h}$ =
0.70, power spectrum normalization $\sigma_8$ = 0.82, tilt $n=0.95$
\citep{Klypin:2011a}. These parameter values are consistent with the
Wilkinson Microwave Anisotropy Probe (WMAP) 5/7-year results
\citep{Komatsu:2009a,Komatsu:2011a}.

When two dark matter halos merge we define the `central' galaxy as the
most massive galaxy of the larger halo, with all other galaxies termed
`satellites.'  The satellite galaxies then lose angular momentum due
to dynamical friction and merge with the central galaxy on a timescale
that we estimate based on fitting functions based on numerical
simulations and provided by \cite{Boylan-Kolchin:2008a}.  Satellites
are tidally stripped during this process, so that satellites with long
merger timescales may become tidally disrupted and destroyed before
they merge with the central galaxy.  In this case, the stars from the
satellite are added to a diffuse stellar halo.  We do not allow
satellite galaxies to merge with other satellites.

\subsection{Gas cooling}
Gas cools and condenses from a reservoir of hot gas in the dark matter
halo.  Before reionization, the amount of hot gas in the halo is equal
to the baryon fraction multiplied by the halo mass; once reionization
begins, the amount of collapsed gas is a function of the timescale of
reionization and the halo mass.  We treat this quantity using a
parameterization from numerical hydrodynamic simulations
\citep{Gnedin:2000a,Kravtsov:2004a}, using $z_{\rm overlap}$ = 12 as the
redshift at which H$_{\rm II}$ regions first overlap and $z_{\rm reion}$ =
11 as the redshift at which the universe is fully reionized.

Our cooling model is similar to the model originally proposed by
  \citet{White:1991a}.  When a halo first collapses, the hot gas is
initialized at the virial temperature of the halo and follows an
isothermal density profile, $\rho_{\hot}(r) = m_{\hot}/(4 \pi
r^{2}\rvir)$, where $m_{\hot}$ is the mass of hot gas and $r_{\rm
  vir}$ is the virial radius of the halo.  This gas then cools from
the center to progressively larger radii on a timescale $t_{\cool}$
dependent on the density of the hot gas.  We can thus define a cooling
radius $r_{\cool}$ as the radius within which all the enclosed gas has
had enough time to cool.  This radius is calculated using the
temperature- and metallicity-dependent atomic cooling curves of
\cite{Sutherland:1993a}.

By setting the cooling time to be equivalent to the dynamical time of
the halo, we find that $t_{\cool} \propto \rvir/\vvir$, where $\vvir$
is the virial velocity of the halo.  If the cooling radius is less
than the virial radius, then solving for the mass within the cooling
radius and differentiating yields a cooling rate of
\begin{equation}
\dot{m}_{\cool}=0.5 m_{\hot}\frac{r_{\cool}} {\rvir}\frac{1}{t_{\cool}}.
\end{equation}
If the cooling rate is larger than the virial radius then the cooling
rate is given by the rate at which gas is accreted into the halo,
\begin{equation}
\dot{m}_{\cool}=0.5 m_{\hot}\frac{1}{t_{\cool}},
\end{equation}
where the factor of 0.5 is included for continuity. 
These two modes of gas accretion are sometimes termed `hot' and `cold' mode
accretion, respectively.  In the `hot' mode gas is assumed to be
shock-heated to the virial temperature of the halo, resulting in
relatively long cooling times.  In the `cold' mode the gas is thought
to penetrate the halo via filamentary streams or `cold flows'
\citep{Birnboim:2003a,Keres:2005a,Dekel:2009a}, and is never
shock-heated.  The transition between these two regimes is dependent
on halo mass and redshift, with cold flows becoming more dominant at
higher redshifts and lower halo masses.

In the SAM we associate cold gas with the central halo, so that only
the central galaxy may accrete gas.  Accreted satellite galaxies
retain their cold gas upon their accretion into a larger halo, but
this gas is typically consumed on a short timescale.  This produces a
population of satellite galaxies that is unrealistically red and old,
as in reality, satellite galaxies may preserve some of their hot gas
haloes, producing a supply of cold gas even after they are
accreted. However, models that have included longer timescale
  accretion onto satellite galaxies have not reported large
  differences in the properties of central galaxies or global
  populations \citep[e.g.][]{Font:2008a,Guo:2011a}, so for the massive
  galaxies that we focus on here, this probably will not have a major
  impact on our results.

\subsection{Disk formation}
As gas cools, we assume that it settles into a rotationally-supported
exponential disk.  Assuming the halo follows an NFW profile
\citep{Navarro:1997a} and responds adiabatically to disk formation, we
can use conservation of angular momentum to find the scale radius of
the disk given the halo's concentration and spin, and the ratio of
baryons in the disk to the mass of the halo.  We use a prescription
based on the work of \cite{Mo:1998a} and described in further detail
in \cite{Somerville:2008b}, which has been shown to reproduce the
scaling relations for disk galaxies out to z $\sim$ 2.  However, in
contrast to \cite{Somerville:2008b}, here we use the baryon fraction
of the disk as predicted by the SAM, and assume that the predicted
scale radius is the scale length of the \emph{gas} (rather than stars)
in the disk.  We convert between the scale lengths of the gas and
stars using $r_{\gas} = \chi \ r_{\rm stars}$, where $\chi$ = 1.7,
based on observations of nearby spiral galaxies
\citep{Leroy:2008}. We have verified that our new models retain good
agreement with the size-mass relation for disk-dominated galaxies, and
its evolution since $z\sim 2$.

\subsection{Disk mode star formation}
We allow for two modes of star formation: `disk mode' star formation,
which occurs in disks at all times as long as cold gas above a
critical surface density is present, and `burst mode' star formation,
which occurs after two galaxies merge or (optionally) after a disk
instability.  Stars are assumed to form following a
\cite{Chabrier:2003a} initial mass function (IMF).  We use an
instantaneous recycling approximation to incorporate stellar mass
loss; at every time-step a fraction $R = 0.43$ of the mass turned into
stars is returned to the cold gas reservoir.  This parameter has been
shown to be a good approximation to the mass loss from massive stars
in a Chabrier IMF \citep{Bruzual:2003b}.

In the `disk mode' the star formation rate density is dependent on the
surface density of cold gas in the disk, following the empirical
Schmidt-Kennicutt relation \citep{Kennicutt:1988a,Kennicutt:1998a}.
Only gas that is above a critical surface density threshold
$\Sigma_{\rm crit} = 6\, \msun\, {\rm pc}^{-2}$ is allowed to form
stars.  All stars that form in the disk mode are added to the disk
component of the galaxy.
 
\subsection{Mergers}
``Wet'' mergers are assumed to trigger a burst of star formation, with
an efficiency that is dependent on the gas fraction of the central
galaxy's disk and the mass ratio of the two progenitors.  This
efficiency is parameterized based on the results of hydrodynamical
simulations \citep{Robertson:2006a,Cox:2008b,Hopkins:2009d}; a higher
efficiency produces a higher star formation rate and destroys a higher
fraction of the disk, transferring that stellar mass to a spheroid.
Following \cite{Hopkins:2009d, Hopkins:2009b} the burst efficiency
decreases with both mass ratio and gas fraction; in extremely gas-rich
disks there is not enough stellar mass to create a torque between the
gas and stars, and so the gas cannot efficiently lose its angular
momentum and collapse.  This is the same approximation as used in
\cite{Somerville:2012a} but represents an improvement relative to
earlier works \citep{Somerville:2008b}, in which the merger efficiency
was strictly a function of mass ratio.

All the stars formed in the `burst' mode are added to the spheroidal
component of the remnant.  We also allow for a fraction $f_{\rm
  scatter} = 0.2 $ of the stars in the satellite galaxy to be
scattered into the diffuse stellar halo; all of the other stars in the
satellite galaxy are added to the bulge.  In addition, we allow a
portion of the central galaxy's stellar disk at final coalescence to
be heated and become dispersion-supported.  Again following
\cite{Hopkins:2009d,Hopkins:2009b} we set the fraction of the disk
that is transferred to the spheroidal component to be equivalent to
the mass ratio of the merger, which is roughly equivalent to the
fraction of the central galaxy's stellar disk that lies within the
radius of the satellite galaxy at coalescence.

We note that, in addition to mergers, environmental processes such as
tidal harassment \citep{Moore:1996a,Moore:1998a,Gnedin:2003a} and ram
pressure stripping of cold gas \citep{Gunn:1972a} have been
shown to drive disks towards earlier morphological types.  We do
  not include any of these processes in our current model. However
these processes are primarily expected to be significant in galaxy
groups and clusters and should be sub-dominant in field galaxy
samples, on which we focus here.
 
\subsection{Supernova feedback and chemical enrichment}
Massive stars and supernovae are assumed to produce winds that drive
cold gas back into the ICM and IGM, heating the gas in the process.
The mass outflow rate is proportional to the star formation rate,
\begin{equation}
\dot{m}_{\rm rh}=\epsilon_{\rm SN}\left(\frac{200\ \rm km \ s^{-1}}{V_{\rm disk}}\right)^{\alpha_{rh}}\dot{m}_{*},
\end{equation}
where $\epsilon_{\rm SN}$ = 1.5 and $\alpha_{rh}$ = 2.2 are free
parameters tuned to reproduce the slope of the stellar mass function
at low masses and $V_{\rm disk}$ is the circular velocity of the disk,
assumed to be the maximum rotation velocity of the dark matter halo
(with this assumption, our disks lie on the observed Tully-Fisher
relation).  The proportion of the gas that is ejected from the halo
entirely is a decreasing function of the halo's virial velocity.  This
gas can then fall back into the hot halo, at a re-infall rate that is
proportional to the mass of the ejected gas and inversely proportional
to the dynamical time of the halo (see S08 for details).

We model chemical enrichment using the instantaneous recycling
approximation.  Whenever a mass d$m_{*}$ of stars is formed, we add a
mass of metals d$M_{\rm Z}\ =y\ dm_{*}$ to the cold gas, where the
chemical yield value y\ =\ 1.5$Z_{\odot}$ is a free parameter that is
chosen to match the normalization of the observed mass-metallicity
relation.  Newly-formed stars are assumed to have the same metallicity
as the mean metallicity $Z_{\rm cold}$ of the cold gas in the ISM at
that timestep.  When gas is ejected due to supernova feedback, these
winds are assumed to have a metallicity $Z_{\rm cold}$. Ejected metals
are assumed to ``re-infall'' back into the hot halo on the same
timescale as the gas.

\subsection{Disk instabilities}
\label{sec:DI}
It is well known that a massive dark matter halo can stablize a
fragile cold, thin disk \citep{OP:73,fall_efstathiou:80}.  Early
dissipationless N-body simulations of isolated disk galaxies
\citep{Efstathiou:1982a} showed that if the mass of stars in the disk
exceeded a critical value relative to the mass of dark matter, the
disk could become unstable and form a bar. Bars can buckle and form
bulges \citep{debattista:04}.

Several previous SAMs have adopted a ``disk instability'' mode for
bulge formation and growth, based on a Toomre-like
\citep{Toomre:1964a} condition to determine the onset of instability
as suggested by \citet{Efstathiou:1982a}. We define 
\begin{equation}
\epsilon_{\rm disk} =\frac{V_{\rm max}}{(G M_{\rm disk}/ r_{\rm disk})^{1/2}},
\label{eqn:epsilon_disk}
\end{equation}
where $V_{\rm max}$ is the maximum circular velocity of the halo,
$r_{\rm disk}$ is the scale length of the disk, and $M_{\rm disk}$ is
the mass of the disk (note we discuss more precise definitions of
these quantities in a moment). Then in any timestep in which
$\epsilon_{\rm disk}< \epsilon_{\rm crit}$, the disk is declared to be
`unstable'. Numerical simulations of idealized, isolated disks suggest
that the value of $\epsilon_{\rm crit}$ is in the range of 0.6-1.1,
with disks containing cold gas having a lower instability threshold then
pure stellar disks
\citep{Efstathiou:1982a,Christodoulou:1995a,Mo:1998a,Syer:1999b}.

Different modelers diverge rather dramatically in deciding the details
of how this criterion is implemented and what the consequences of a
declaration of instability should be. Some modelers use the stellar
mass of the disk as $M_{\rm disk}$ in the formula above, while others
use the sum of the stars and cold gas in the disk. When the disk is
declared to be unstable, some \citep[e.g.][]{Guo:2011a} move just
enough stars from the disk component to the spheroid component to
achieve marginal stability; others \citep[e.g.][]{Bower:2006a} move
\emph{all} of the stars and cold gas in the disk to a bulge
component. Unsurprisingly, these workers differ significantly in their
conclusions regarding the importance of disk instabilities for bulge
growth. Moreover, \citet{Athanassoula:2008a} has argued that this
criterion is in any case insufficient to determine the onset of bar
formation.

In addition to this ``classical'', bar-driven mode of in-situ bulge
growth, cosmological hydrodynamical simulations
\citep[e.g.][]{Dekel:2009b,Ceverino:2010a} have shown that gas-rich
disks can undergo dramatic fragmentation due to ``violent disk
instabilities'' (VDI). In VDI, gravitational instabilities lead to the
formation of giant clumps in the disk, which appear qualitatively
similar to the structures that are frequently observed in high
redshift disks
\citep{elmegreen:2009,Genzel:2011a,y.guo:12,Wuyts:2012b}. These clumps
may then migrate to the center of the galaxy, providing an alternative
pathway for bulge growth
\citep{elmegreen:2008,Dekel:2009b,Bournaud:2011a}. Disks are more
susceptible to VDI at high redshift, where the accretion rate and gas
fractions are high \citep{Dekel:2009b}.  While there has been recent
progress in studying this process, the amount of mass actually
transferred to the bulge is highly sensitive to the effectiveness of
stellar-driven winds. In simulations with strong stellar feedback,
many of the clumps tend to be destroyed before they reach the center
of the galaxy \citep{Genel:2012c}, so they probably contribute little
to the bulge. However, the implementation of these feedback processes
in galaxy-scale simulations remains highly uncertain.

Keeping in mind these large uncertainties, we consider three ways of
treating disk instabilities in this paper. In the first ({\bf `No
  DI'}), we neglect disk instabilities and allow bulges to form and
grow only through mergers. This case corresponds to most previously
published versions of the Santa Cruz SAM (e.g., S08, S12). In our
second model variant ({\bf `Stars DI'}), we use the radius and mass of
the stellar disk as $r_{\rm disk}$ and $M_{\rm disk}$ in
Eqn.~\ref{eqn:epsilon_disk} above, and in each timestep in which the
disk is deemed to be unstable, we move \emph{stars only} from the disk
to the spheroid component until $\epsilon_{\rm disk} = \epsilon_{\rm
  crit}$. In our third model variant ({\bf `Stars+Gas DI'}), we use
the radius of the \emph{gas disk} and the \emph{mass of stars and cold
  gas} in the disk as $r_{\rm disk}$ and $M_{\rm disk}$ in
Eqn.~\ref{eqn:epsilon_disk} above. In each timestep in which the disk
is deemed to be unstable, we move \emph{stars and gas} from the disk
to the spheroid component to achieve marginal stability, and the cold
gas that is moved to the spheroid is assumed to participate in a
starburst. If we define the total mass of gas and stars that is moved
from the disk to the bulge to be $\Delta m_{\rm bulge}$, then the mass
of cold gas that is moved to the bulge is $\Delta m_{\rm bulge, gas} =
f_{\rm cold, disk} \Delta m_{\rm bulge}$, where $f_{\rm cold, disk}
\equiv m_{\rm cold, disk}/(m_{\rm cold, disk}+m_{\rm star, disk})$ is
the cold gas fraction in the disk. The efficiency and timescale of the
starburst is computed using the same approach as for merger-triggered
bursts. In both models, we use the maximum circular velocity of the
(unmodified) dark matter halo as $V_{\rm max}$ in Eqn.~\ref{eqn:epsilon_disk}.

Our implementation of the `Stars DI' model is very similar to that
used by other authors such as \citet{De-Lucia:2011a} and
\citet{Guo:2011a}. Our `Stars+Gas DI' model is intended to
schematically represent a VDI-like situation, which is driven largely
by the inflow of cold gas. In practice, because we restore the disk to
marginal stability in each timestep in which it is determined to be
unstable, especially at high redshift the disk tends to become
unstable again in the next timestep due to accretion of new gas which
grows the disk above the critical mass again, producing a cascade of
instability events in the SAM.

In agreement with some previous studies, we find that our `No DI'
model does not appear to produce enough intermediate mass early-type
galaxies at $z=0$, and that this agreement is improved by switching on
the disk instability mode for bulge growth
\citep{Guo:2011a,De-Lucia:2011a,Shankar:2013a}. We therefore tune our
model by adjusting $\epsilon_{\rm crit}$ to reproduce the observed
stellar mass function of early-type galaxies in the local universe. We
adopted $\epsilon_{\rm crit}$ = 0.75 in the `Stars DI' model and
0.70 in the `Stars+Gas DI' model, which is within the expected
range of values suggested by the simulations discussed above.

\subsection{Black Hole Accretion}

Each galaxy is seeded with a `heavy' black hole
\citep{Loeb:1994a,Koushiappas:2004a,Volonteri:2011a}, with a mass
$10^{5} \ \msun$. In all of our models, as in S08, rapid black hole
accretion leading to radiatively efficient AGN activity is triggered
by mergers. When two galaxies merge, their black holes are assumed to
merge as well. As the two galaxies approach coalescence the black hole
begins to accrete and radiate, depositing energy into the ISM. The
black hole accretes at the Eddington rate until it exceeds a critical
mass; this mass corresponds to the energy needed to halt further
accretion and begin to power a pressure-driven outflow. The accretion
rate then declines as a power law, following the lightcurves derived
from numerical simulations by \citet{hopkins:06}. For more details,
see \citet{Hirschmann:2012a}. 

Following \cite{Hopkins:2007a}, the critical mass is given by
\begin{equation}
\log\left(\frac{M_{\rm crit}}{M_{*,\rm bulge}}\right)=f_{\rm BH, crit} \, [-3.27+0.36 \erf((f_{\rm gas}-0.4)/0.28)],
\label{eqn:Mcrit}
\end{equation}
where $M_{\rm *,bulge}$ is the stellar mass of the bulge, $f_{\rm
  gas}$ is the cold gas fraction of the larger merger progenitor, and
$f_{\rm BH, crit}$ is a tunable parameter of order unity, set to
reproduce the redshift zero relationship between the mass of the black
hole and the mass of the stellar bulge \citep{McConnell:2013a}. This
fitting function is based on an analysis of a large suite of SPH
simulations of binary mergers including BH growth and AGN feedback, as
described in \citet{Hopkins:2007a}.  If the sum of the two
pre-existing black holes is already larger than the critical mass,
then the BH goes immediately into the ``blowout'' phase (i.e., the
accretion rate starts at Eddington but immediately begins to
decline). Note that this is slightly different from the implementation
in S08 and \citet{Hirschmann:2012a}, in which if the initial BH mass
was larger than the critical mass, no accretion onto the BH was
triggered at all. The new treatment is a more realistic representation
of what actually happens in the hydrodynamic simulations (P. Hopkins,
priv. comm.).

In the `Stars DI' and `Stars+Gas DI' models, black holes are also
allowed to grow following a disk instability. As with mergers, the
growth of the black hole is limited by the amount of low angular
momentum material in the center of the galaxy following an instability
event.  We limit this term to be a fraction $f_{\rm fuel,DI} = 0.002$
of the mass that is transferred from the disk to the bulge. Following
a DI event, the BH accretes until this fuel is consumed. However, as
noted above, the high gas inflow rate at high redshifts tends to
produce gas-rich disks that are unstable for extended periods of time
in the `Stars+Gas DI' model.  Allowing the black hole to accrete at
the Eddington rate whenever the disks are unstable would produce black
holes that are too massive at high redshifts, and too many luminous
AGN. Therefore, following \citet{Hirschmann:2012a}, we limit the
DI-triggered black hole accretion rate to be a fraction of the
Eddington limit. We choose this fraction to have a mean of $f_{\rm
  Edd,DI} = 0.01$, and to have a Gaussian scatter from timestep to
timestep of 0.2, representing the likely stochasticity of the BH
accretion. This choice reproduces the AGN luminosity function from
$z\sim 6$--0 (Hirschmann et al. in prep) and also still reproduces the
BH mass bulge mass relation at $z=0$. However, bulges and also BH form
earlier in the `Stars+Gas DI' model, so we must also decrease $f_{\rm
  BH, crit}$ to 0.6 in the merger triggered BH accretion mode
(Eqn. \ref{eqn:Mcrit}) to maintain the agreement with the observed
$m_{\rm BH}-m_{\rm bulge}$ relation.
  
Black holes are also allowed to grow via Bondi-Hoyle accretion
\citep{Bondi:1952a} from the hot halo. This typically very
sub-Eddington mode of accretion is associated with heating of the hot
gas via giant radio jets, in what is sometimes called `radio mode'
feedback. The strength of the `radio mode' feedback is governed by a
tunable parameter $\kappa_{\rm radio}$ (see S08 for the precise
definition), which we tuned independently in all three models to
reproduce the knee and high-mass end of the stellar mass function.

\section{Model for structural parameters of spheroids}
\subsection {Effective radius}

We first consider spheroids that are formed in mergers. In the case of
a merger without dissipation, simple conservation of energy arguments
would predict that the internal energy of the two progenitors is
conserved during the merger:

\begin{equation}
E_{\rm init} = E_{\rm f} = 
		\Cint\sum_{i=1}^{2}G\frac{(M_{*,i}+M_{\rm new*,i})^{2}}{R_{*,i}}=\Cint G\frac{M_{*, \rm f}^{2}}{R_{\rm *, f}},
		\label{eqn:Energy}
\end{equation}
where $M_{*,i}$ is the stellar mass of each of the two progenitors,
$M_{\rm new*,i} $ is the mass of stars formed during the merger,
$R_{*,i}$ are the three dimensional effective radii of the
progenitors, $M_{\rm *, f}$ and $R_{\rm *, f}$ are the stellar mass and
3D effective radius of the merger remnant, and $\Cint$ is a dimensionless
constant relating the internal energy of the galaxy to $GM^{2}/R$.
However, in the presence of gas, mergers can be highly dissipative,
inducing large amounts of star formation
\citep{Dekel:2006a,Robertson:2006b,Hopkins:2009b}; thus the
conservation of energy relation must be modified with a term
incorporating radiative losses.  We note that the parameter $\Cint$ in
equation \ref{eqn:Energy} may actually have a degree of dependence on
the morphology of the galaxy, but we have not attempted to account for
that here.

Motivated by the results of hydrodynamical binary merger
simulations \citep[2008]{Cox:2006b}, \cite{Covington:2008b} provided a
simple parameterization of this radiative energy loss:
\begin{equation}
		\label{eqn:Energy_radiated}
		E_{\mathrm{rad}}=\Crad\sum_{i=1}^{2}K_{i}f_{\mathrm{g},i}f_{\mathrm{k},i}(1+f_{\mathrm{k},i}),
	\end{equation}
where $K_{\emph{i}}$, $f_{\mathrm{g},i}$, and $f_{\mathrm{k},i}$ are
the total kinetic energy, baryonic gas fraction, and fractional
impulse of progenitor \emph{i}, $\Crad$ is a dimensionless constant
and the sum is over the two progenitors.  Adding this term to the left
hand side of equation \ref{eqn:Energy} provides a natural way to
incorporate dissipation in the calculation of the effective radius of
elliptical galaxies ($E_{\rm init} + E_{\rm rad} = E_{\rm f}$).
Mergers with higher amounts of dissipation and star formation produce
remnants with smaller effective radii, allowing for the creation of
compact elliptical galaxies from diffuse spiral galaxies.

These two formulae have previously been shown to provide accurate
predictions for the effective radii of elliptical galaxies resulting
from the gas-rich mergers of spiral galaxies
\citep[2011]{Covington:2008b}.  In this case, the constants $\Crad $
and $\Cint $ were measured by fitting the relations to a suite of
hydrodynamical merger simulations \citep[2008]{Cox:2006b}.
Incorporating a simplified version of these formulae in semi-analytic
models has been shown to reproduce the size-mass relation of
early-type galaxies \citep{Shankar:2013a,Covington:2011a}.

The model presented in \citet{Covington:2008b} and
\citet{Covington:2011a} was limited in that it was only calibrated
against simulations of fairly gas-rich mergers of disk-dominated
progenitors. Here, we wish to implement the model self-consistently
within a full SAM, in which mergers between galaxies with a wide
variety of initial gas fractions and morphologies are predicted to
occur. Therefore we extend the model using an additional 68
hydrodynamical simulations of binary mergers, described in
\cite{Johansson:2009a}, and carried out using the GADGET-2 Smoothed
Particle Hydrodynamics (SPH) code \citep{Springel:2005b}. The
simulations adopt the sub-grid ISM model of \citet{Springel:2003a},
and the \citet{Springel:2005d} treatment of BH growth and AGN feedback
via thermal heating. A strongly pressurized ``stiff'' Effective
Equation of State is adopted, to allow for construction of stable gas
rich disks, following \citet{Springel:2005b}. Stellar driven winds are
not included. These ingredients are similar to those adopted in the
Cox et al. simulations used in the analysis of Covington et al., but
with the addition of BH growth and AGN feedback.

This suite includes major (mass ratio $>$ 3:1) and minor mergers
between two disk-dominated galaxies (D-D, D-d), and between a
disk-dominated and spheroid-dominated galaxy (B-D, B-d).  It also
contains major mergers between two spheroid-dominated galaxies (B-B);
the only permutations lacking are minor mergers between two
spheroid-dominated galaxies (B-b), and mergers between a larger
disk-dominated galaxy and a smaller spheroid-dominated galaxy (D-b).

The spheroid-dominated progenitor galaxies were formed from mergers of
disk-dominated galaxies with gas fractions ranging from 20\% to 80\%,
producing a range of stellar masses and gas fractions in all of the
mergers.  We have measured the three-dimensional half-mass effective
radii of these merger remnants directly, and used these values to perform
a $\chi^{2}$ fit to constrain $\Cint$ and $\Crad$.  We calculate
$\Cint$ and $\Crad$ independently for the five categories of mergers
described above.

\begin{figure*}
	\begin{center}
		\includegraphics[width=0.7\textwidth]{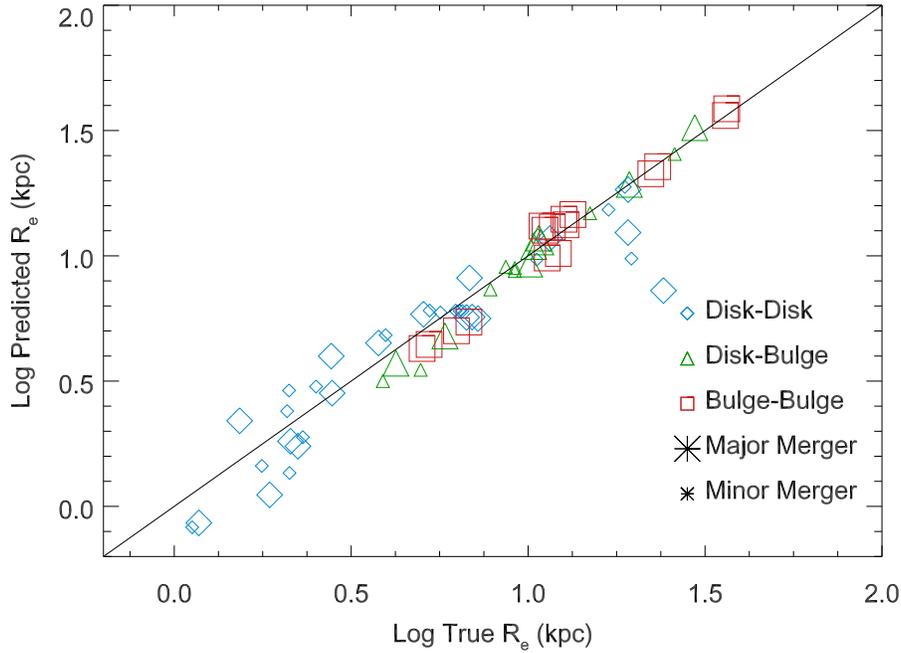}
		\caption{Predicted versus measured effective radius
                  for the merger remnants of
                  \protect\cite{Johansson:2009a}.  The constants
                  $\Cint$ and $\Crad$ were calibrated independently
                  for major (large symbols) and minor (small symbols)
                  mergers, depending on the morphology of the
                  progenitors (see Table \protect\ref{Constants}).
                  Blue diamonds represent mergers between two
                  disk-dominated galaxies, green triangles represent
                  mergers between a disk-dominated and a
                  spheroid-dominated galaxy, and red squares represent
                  mergers between two spheroid-dominated galaxies.}
		\label{fig:Radius}
	\end{center}
\end{figure*}

Results are shown in Table \ref{Constants} and Figure
\ref{fig:Radius}.  The value of $f_{\rm rad}\equiv\Crad/\Cint$ can be
thought of as characterizing the relative importance of dissipation;
high values indicate more dissipation.  We find that this value is
highest for major mergers of two disk-dominated galaxies ($f_{\rm rad}
= 5.0$), is lower for minor mergers between two disk-dominated
galaxies ($f_{\rm rad} = 2.7$) and is zero for mergers where one or
both of the galaxies is spheroid-dominated.  This latter subset of
mergers is thus essentially dissipationless; considering the average
baryonic gas fraction in this group of mergers is 5.5\% this is not an
unexpected result.  We note that the largest disk-disk merger remnants
are much larger than predicted; these galaxies had high gas fractions
before the mergers, and large amounts of star formation during the
mergers.  Thus these outliers may be an indication that the model
breaks down in the most extreme merger events.

\begin{table}
	\begin{center}
		\begin{tabular}{lllr}
			\hline
			Merger & $\Cint$ & $\Crad$ & Number\\
			\hline
			D-D  & 0.50 & 2.50  & 18\\
			D-d  & 0.50 &  1.35 & 18\\
			B-D  & 1.00  & 0.00  & 11\\
			B-d  & 1.00  & 0.00 & 8\\
			B-B & 1.00 & 0.00 & 11\\
			\hline
		\end{tabular}
	\end{center}
	\caption{Best-fit calculations of $\Cint$ and $\Crad$ for the
          \protect\cite{Johansson:2009a} hydrodynamical simulations.
          Progenitor types B and D designate bulge- and disk-dominated
          galaxies, respectively.  Capital (lowercase) letters denote
          the larger (smaller) galaxy; capital-capital pairs indicate
          a major merger, capital-lowercase pairs indicate a minor
          merger.  Lower values of $\Crad/ \Cint$ represent mergers
          with less dissipation; mergers where $\Crad$ = 0.0 are
          essentially dissipationless.}
	\label{Constants}
\end{table}

\subsection{Velocity dispersion}
We use the virial theorem to determine the line-of-sight velocity
dispersion of the remnant:
	\begin{equation}
		\label{eqn:velocity_dispersion} 
		\sigma^2=\left(\frac{\Csig G}{2R_{\rm f}}\frac{M_{\rm *,f}}{(1-f_{\rm{dm,f}})}\right),
	\end{equation} 
where $M_{\mathrm{*,f}}$ is the stellar mass of the remnant (or the
single galaxy, in the case of disk instabilities), $R_{\mathrm{f}}$ is
the stellar half-mass radius of the remnant, and $\Csig$ is a
dimensionless constant that accounts for the conversion between the
three-dimensional effective radius and the line-of-sight projection of
the velocity dispersion.  We define $M_{\rm dm}$ to be the mass of
dark matter within $R_{\rm f}$, and $f_{\mathrm{dm,f}} = M_{\rm
  dm}/(0.5M_{\rm *,f}+M_{\rm dm})$ to be the central dark matter
fraction of the remnant (i.e. the proportion of mass within the
stellar effective radius that is dark matter).  Thus the term $0.5
M_{\rm *,f} +M_{\rm dm} = 0.5M_{\rm *,f}/(1-f_{\rm{dm,f}})$ represents
the total amount of stars and dark matter within the stellar half-mass
radius.

\begin{figure*}
	\begin{center}
		\includegraphics[width=0.7\textwidth]{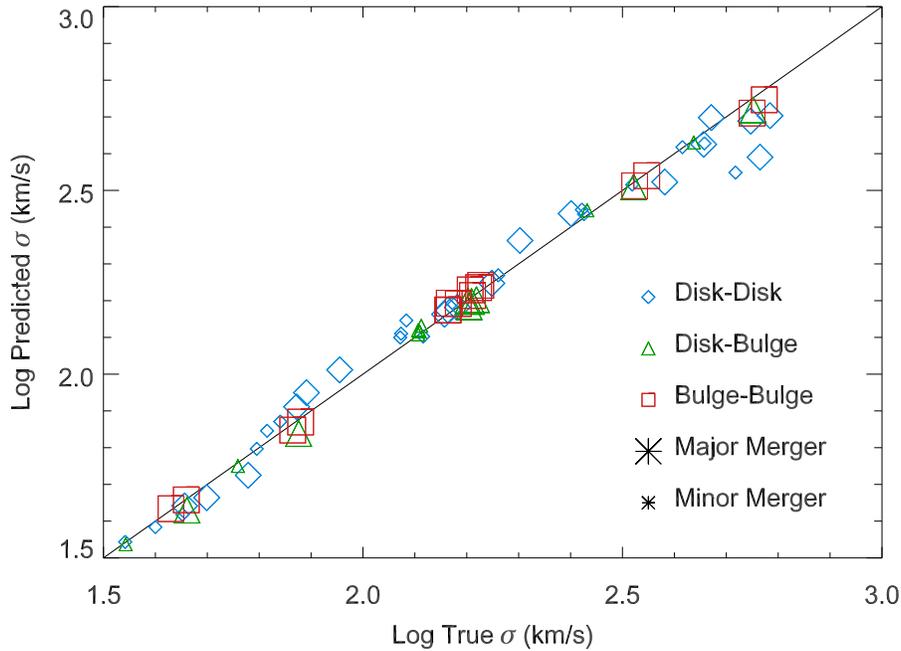}
		\caption{Predicted versus measured velocity dispersion
                  for the merger remnants of
                  \protect\cite{Johansson:2009a}.  The constant
                  $\Csig$ was calibrated independently for major
                  (large symbols) and minor (small symbols) mergers,
                  depending on the morphology of the progenitors.  All
                  calibrations resulted in values $0.29 < \Csig <
                  0.31$; we adopt $\Csig = 0.30$ for all mergers. Blue
                  diamonds represent mergers between two
                  disk-dominated galaxies, green triangles represent
                  mergers between a disk-dominated and a
                  spheroid-dominated galaxy, and red squares represent
                  mergers between two spheroid-dominated galaxies.}
		\label{fig:Sigma}
	\end{center}
\end{figure*}

We have calculated $\Csig$ using a least-squares fit for the five
categories of merger simulations described above.  Stellar velocity
dispersions were measured within the half-mass radius, using the
average of 50 random line-of-sight projections.  In this fit, the
`true' (rather than the `predicted') half-mass radius from the
simulations was used.  In all cases, the value of $\Csig$ was between
0.29 and 0.31; thus we adopt the value $\Csig = 0.30$ for all mergers.
Results are shown in Figure \ref{fig:Sigma}.  We find that this model
quite accurately reproduces the velocity dispersions of merger
remnants.

\subsection{Implementation within the SAM}
We apply these prescriptions in the SAM whenever two galaxies with a
mass ratio greater than 1:10 merge (in mergers more minor than 1:10,
the satellite material is added to the disk component as indicated by
simulations). As the hydrodynamical merger suite does not contain
minor mergers below a 1:6 mass ratio, we note that the prescription
for effective radius and velocity dispersion has not been tested for
mergers between 1:6 and 1:10, which are actually very common.  
For the two merger cases not covered by our simulation suite,
  spheroid-spheroid minor mergers (B-b) and disk-spheroid minor
  mergers between a larger disk and a smaller spheroid (D-b), we adopt
  $C_{\rm rad} = 0.0$, $C_{\rm int}=1.0$, and $C_{\rm rad} = 1.35$,
  $C_{\rm int}=0.5$ respectively. We expect that the morphology of the
  secondary galaxy will not have a major effect on the remnant in
  minor mergers.

We assume orbital parameters for the merger following a statistical
distribution \citep{Wetzel:2010a} that is dependent on the redshift of
the merger and the mass of the halo containing the more massive
progenitor.  Whereas in the hydrodynamical simulations we were able to
measure the contribution of dark matter within one effective radius
directly, in the SAM we assume that the two dark matter halos merge
dissipationlessly and that the remnant halo follows an isothermal
profile.  Strong lensing studies suggest this to be a reasonably
accurate model for massive early-type galaxies \citep{Auger:2010a}.

The above formulae have only been calibrated against hydrodynamical
simulations of binary mergers, but in the SAM we may also form a
stellar bulge through disk instabilities.  In this case we follow the
prescription of \cite{Guo:2011a}, assuming that the new bulge mass
forms from the center of the stellar disk, which has an exponential
surface density profile: $\Sigma(r) =\Sigma_{0} \exp(-r/r_{\rm d})$,
$\Sigma_{0}=M_{d}/2\pi r_{\rm d}^{2}$, where $M_{\rm d}$ is the mass
of the disk and $r_{\rm d}$ is the scale length of the stellar
disk. The radius $r_{\rm pb}$ of the disk enclosing this proto-bulge
is then found by solving:
\begin{equation}
		M_{\rm pb}=\frac{M_{\rm d}}{r_{\rm d}}\left[r_{\rm d}-e^\frac{-r_{\rm pb}}{r_{\rm d}}(r_{\rm d}+r_{\rm pb})\right],
		\label{eqn:Diskinstability}
\end{equation}
for $r_{\rm pb}$ where $M_{\rm pb}$ is the mass of the disk transferred
to the bulge in the instability event.  We then assume that this
stellar proto-bulge merges dissipationlessly with any existing bulge.

Having calculated the radius of the bulge following the disk
instability, we model velocity dispersion using the same extension of
the virial theorem that we use for mergers.

We compute an effective radius for the composite bulge plus disk
system by calculating a stellar mass- or light-averaged radius:
\begin{equation}
r_{\rm eff} = (r_{\rm eff, d} M_{\rm d} + r_{\rm eff, b} M_{\rm b})/(M_{\rm d}+M_{\rm b})
\end{equation}
where $r_{\rm eff, d} = 1.67\, r_{\rm d}$ is the effective radius of
the disk, $r_{\rm eff, b}$ is the effective radius of the spheroid,
and $M_{\rm d}$ and $M_{\rm b}$ are the stellar mass or luminosity of
the disk and spheroid, respectively. We have found that this simple
approach agrees well with more detailed modeling of the effective
radius of composite spheroid plus disk systems. We use a rest-frame
$r$-band luminosity weighting throughout this paper, but find nearly
indistinguishable results when using stellar mass weighting.

We then convert our 3D radii to 2D projected radii, using the same
approach as that suggested by \citet{Shankar:2013a}. This approach
relies on the tabulated form factors computed by
\citet{Prugniel:1997a}. These factors $S(n)$ relate the gravitational
energy $|W|$ to the effective radius $R_e$, for a system of total mass
$M$ and with S\'{e}rsic index $n$:
\begin{equation}
|W| = S(n) \frac{GM^2}{R_e} = \frac{GM^2}{R_g} \, ,
\end{equation}
where $R_g$ is the gravitational radius. 
If the system is virialized, we can adopt the approximation $R_g
\simeq 2r_{\rm eff, 3D}$.  The projected radius is then given by:
\begin{equation}
R_e = 2\, S(n)\, r_{\rm eff, 3D}
\end{equation}
where $r_{\rm eff, 3D}$ is the 3D effective radius. We adopt
$S(n)=0.34$ from \citet{Prugniel:1997a}, appropriate for a S\'{e}rsic
index of $n=4$. However, the dependence on S\'{e}rsic index is weak,
so we adopt this conversion factor for both spheroids and disks.

An additional issue in comparing with observations is that our radii
are estimated in terms of the stellar mass, while most observational
studies have measured the radial extent of the
\emph{light}. \citet{szomoru:12} found that the radius of the stellar
mass distribution in galaxies of all types is on average 20-30\%
smaller than that of the rest-frame $g$-band light, with no strong
apparent trends with galaxy properties or redshift, though their
sample was fairly small. We do not apply this correction here, but we
should keep in mind that the absolute normalization of the
observational size estimates could shift by this much. 

\section{Results}

Because of the uncertainties surrounding the formation of bulges
through disk instabilities discussed above, we present results for the
three different treatments of this process as described in
Section~\ref{sec:DI}: `No DI', `Stars DI', and `Stars+Gas DI'.  In the
`No DI' models, bulges can only grow through mergers, while in the
`Stars DI' model, stars can be moved from the disk to the bulge, and
in the `Stars+Gas DI' model, both stars and gas can be moved from the
disk to the bulge, and gas moved to the bulge participates in a
starburst. In addition, in the `Stars DI' model, the DI criterion is
determined only by the properties of the stellar disk, while in the
`Stars+Gas DI' model, the DI criterion depends on the mass of both
stars and gas in the disk. We tune the above-mentioned parameters for
black hole growth and disk instabilities to reproduce the stellar mass
function and black hole-stellar bulge mass scaling relations at
redshift zero.  In the two DI models we apply the additional
constraint of the fraction of spheroid-dominated galaxies as a
function of stellar mass at $z=0$, to fix the instability parameter
($\epsilon_{\rm crit}$), as we now describe.

\subsection{Model Calibration}
  
\begin{figure}
\centering
\includegraphics[width=0.45\textwidth]{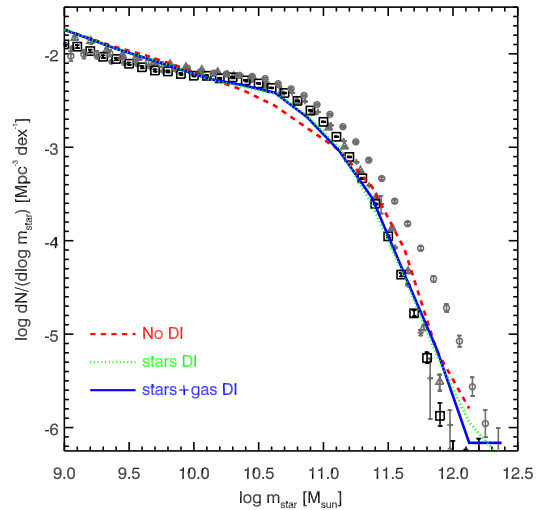}
\caption{The galaxy stellar mass function at redshift zero.  The three
  colored lines show the `No DI' (red), `Stars DI' (green) and
  `Stars+Gas DI' (blue) models. The symbols show observational
  estimates of the stellar mass function of nearby galaxies
  (\protect\cite{Moustakas:2013a}: squares; \protect\cite{li:09}:
  triangles; \protect\cite{panter:07}: crosses;
  \protect\cite{Bernardi:2013a}: circles).  }
\label{fig:MassFunction}
\end{figure}

We first consider the stellar mass function for all galaxies (Figure
\ref{fig:MassFunction}). As discussed in Section 2.7, the efficiency
of the radio mode feedback in quenching star formation is tuned
independently in all three models to reproduce the knee and high-mass
end of the mass function as estimated by \citet{Moustakas:2013a}.
We note that all of the models either fall slightly below the knee of
the observed stellar mass function or lie slightly above the observed
high-mass turnover.  This is an indication of the tension between
matching the turnover and the high-mass cutoff of the mass function;
increasing the strength of the radio mode feedback can provide a
better match to the high-mass end but then underpredicts the knee of
the stellar mass function.
 
However, the observational estimates of the stellar mass function at
the high mass end are highly uncertain. To illustrate this, we also
show the observed stellar mass function of \cite{Bernardi:2013a},
which estimates the total light from the galaxy using a combination of
a S\'ersic and exponential profile.  The results differ both in the
estimated luminosities, due to different choices for how to do the
photometry and background subtraction, and also in the assumed stellar
mass-to-light ratios. We could tune our model to instead match the
\cite{Bernardi:2013a} mass function, simply by reducing the strength
of the AGN feedback, but due to the remaining uncertainties in the
observations, we choose to retain the more conventional normalization
for this work.
 
We now turn to the mass function divided by galaxy morphological
type. The only morphological information provided by our SAM is the
mass or luminosity of stars in the disk and spheroidal components, or
equivalently a bulge-to-total ratio ($B/T$). It is not entirely
straightforward to determine how to compare this with the most robust
and widely available observational morphological
classifications. These include ``Hubble types'' based on visual
classifications, concentration (usually defined as the radius
containing 90 percent of the light divided by that containing 50\%) or
S\'ersic indices based on the light profiles, and bulge-disk
decompositions. Unfortunately, we were unable to locate any published
stellar mass functions divided according to a $B/T$ determined from a
bulge-disk decomposition, which would be the most straightforward to
compare with our predictions. We chose to use observational stellar
mass functions divided by $r$-band concentration ($c_r \equiv
r_{90}/r_{50}$). \citet{gadotti:09} found, based on a sample of 1000
galaxies from the Sloan Digital Sky Survey (SDSS) with bulge-disk
decompositions, that $B/T$ correlates more tightly with $c_r$ than
with S\'ersic index. However, \citet{Cheng:2011a} found in an analysis
of SDSS galaxies that concentration-selected samples can include a
significant population of Sa and S0 galaxies in addition to true
ellipticals. Moreover, many visually classified disks have
$B/T>0.5$. We will use $B/T$ to define spheroid-dominated galaxies,
which we refer to more or less interchangably as ``early types'', but
we should keep in mind the difficulty of selecting the same types of
objects in the observations.

It is also unclear what value of $B/T$ we should use to define our sample
of ``early types''. \citet{Guo:2011a} argue that a $B/T$ of 0.2 should
correspond to a concentration of 2.86, which is a commonly used value
to define ``early type'' galaxies \citep[e.g.][]{Shen:2003a}. Based on
an eyeball fit to the plot of $B/T$ versus concentration shown in
\citet{gadotti:09}, $c_r \simeq 2.6$--2.9 would correspond to $B/T
\simeq 0.4$, though it is clear from the figure that selecting
galaxies via concentration is not equivalent to a clean cut in
$B/T$. \citet{gonzalez:09} predict based on model galaxies taken from
the GALFORM semi-analytic model that galaxies with $c_r = 2.86$ may
have a broad range of values of $B/T$ ranging from about 0.25 to about
0.65. Figure 9 of \citet{Cheng:2011a} shows that a galaxy sample
selected with $c>2.9$ would be expected to contain a significant
population of galaxies with $0.2<B/T<0.5$. Many other studies adopt
higher values of $B/T \sim 0.5$--0.7 to define early types
\citep{Shankar:2013a,wilman:13}.

\begin{figure*}       
\centering
\includegraphics[width=\swidth]{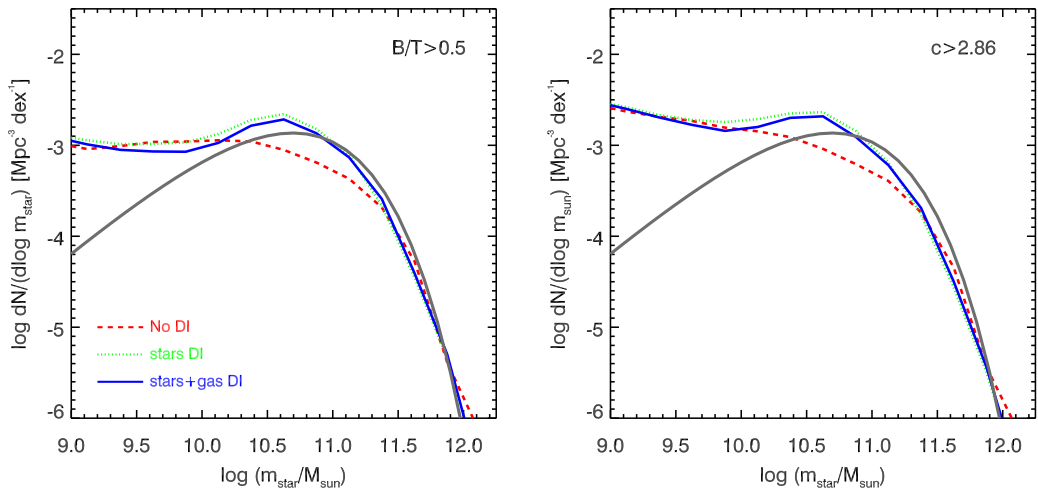}
\caption{Stellar mass function for ``early type'' galaxies at redshift
  zero. In both panels, the gray curves show the observational
  estimate of the stellar mass function for galaxies with
  concentration $c>2.86$ from \protect\cite{bernardi:10}. The three
  colored lines show the `No DI' (red), `Stars DI' (green) and
  `Stars+Gas DI' (blue) models. In the left panel, the models are
  selected according to the stellar mass bulge-to-total ratio
  ($B/T>0.5$). In the right panel, we have assigned concentrations to
  the model galaxies using an empirical scaling (see text), and
  selected them to have $c>2.86$. The model in which bulges form and
  grow only due to mergers (No DI) appears to underproduce early type
  galaxies in the mass range $\sim 2 \times 10^{10}$ to $\sim 3 \times
  10^{11} \msun$.
}
\label{fig:MassFunctionMorph}
\end{figure*}

Figure~\ref{fig:MassFunctionMorph} shows the stellar mass function
predicted by our three models with different recipes for bulge
formation: `No DI', `stars DI' and `stars+gas DI', compared with the
observational estimate of the stellar mass function for
spheroid-dominated galaxies from \citet{bernardi:10}. We compare
galaxies selected with $B/T>0.5$ with the observed mass function for
$c_r = 2.86$, and we also assign concentrations to our model galaxies
using a simple empirial scaling from the results of
\citet{gadotti:09}: $c_r = 2.0 + 2.4 \,B/T$ for $B/T<0.5$; and $c_r =
3.0$ for $B/T>0.5$, plus a Gaussian random deviate with $\sigma =
0.2$. We can see that these two approaches yield similar
results. Moreover, we see that the model in which spheroids can only
grow via mergers has too few intermediate-mass ($10^{10} < \msun <
10^{11}$) early-type galaxies.  A similar result has been found in
other SAMs \citep{Parry:2009a,De-Lucia:2011a}, which use different
prescriptions for spheroid growth in disk instabilities and mergers.
In contrast our `Stars DI' and `Stars+Gas DI' models match the
observed early-type mass function fairly well for galaxies with $m_*
\gtrsim 10^{10} \msun$.

\begin{figure*}       
\centering
\includegraphics[width=\textwidth]{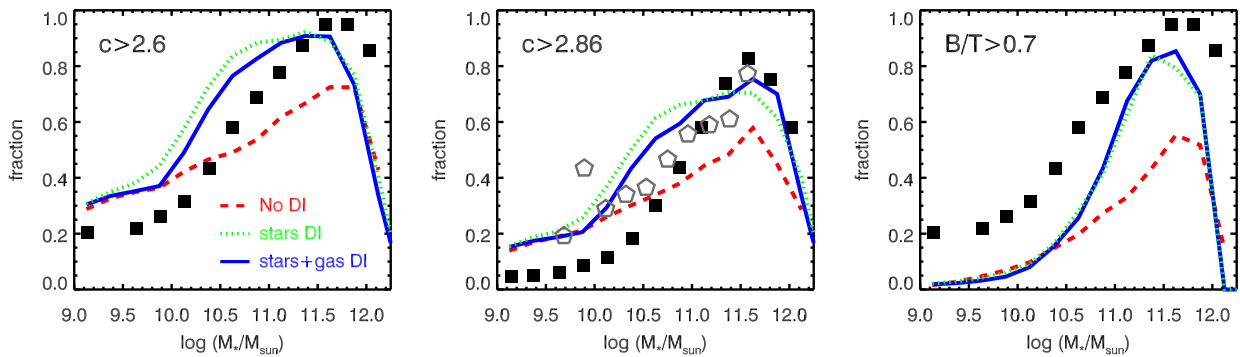}
\caption{The fraction of ``early type'' galaxies at $z=0$. The three
  colored lines show the `No DI' (red), `Stars DI' (green) and
  `Stars+Gas DI' (blue) models.  In the left panel, concentrations
  have been assigned to the model galaxies using an empirical scaling
  (see text), and the fraction of model galaxies with $c>2.6$ is
  shown. The observed fraction of galaxies with $c>2.6$ from
  \protect\cite{bernardi:10} is shown with the black square
  symbols. In the middle panel, model predictions and observational
  results are shown for galaxies with $c>2.86$ in a similar
  manner. Observational estimates from the early-type sample
  classified by \protect\cite{Cheng:2011a} are shown with open
  pentagons; this sample corresponds approximately to $B/T>0.5$ or
  $c>2.9$. In the right panel, we show model galaxies with stellar
  mass bulge-to-total ratio $B/T>0.7$, compared with the early type
  sample of \protect\cite{Hyde:2009a}, shown with filled squares.
Again we see that the model without disk instabilities fails to
produce enough spheroid-dominated galaxies at intermediate masses,
regardless of the criteria used. }
\label{fig:Morphfrac}
\end{figure*}

Figure~\ref{fig:Morphfrac} shows the fraction of spheroid-dominated
galaxies for several other cuts in $B/T$ or concentration, along with
observational estimates. The \cite{Hyde:2009a} sample of early
  type galaxies is selected via the SDSS parameter fracDev$=1$
  (fraction of light well-fit by a de Vaucouleur profile) and $r$-band
  axis ratio $b/a>0.6$; we assume that this corresponds approximately
  to $B/T>0.7$ for our model comparison. The \citet{Cheng:2011a}
  early-type sample is selected using an automated classification
  method based on image smoothness, axial ratio, and concentration and
  tuned to match the results of human visual classification. They
  state that their early-type sample corresponds approximately to
  $B/T>0.5$, $c>2.9$, so we plot it in our $c>2.86$ panel.
In all cases, the `No DI' model significantly underproduces
spheroid-dominated galaxies at intermediate masses, while the models
with bulge formation via disk instabilities perform much better in
this regard (though too many spheroid-dominated galaxies appear to be
produced at low masses).  Although the uncertainties are still
considerable, these results are suggestive that mergers may not be the
only significant channel for building spheroids. We reiterate,
  however, that there are still significant differences in the
  morphological fractions derived from observations using different
  classification methods, and none of the current comparisons are able
  to directly compare the theoretically predicted quantity ($B/T$)
  with observations, making it difficult to obtain precise and robust
  constraints on the models. This is clearly an area where further
  work is required.

\begin{figure}       
\centering
\includegraphics[width=0.45\textwidth]{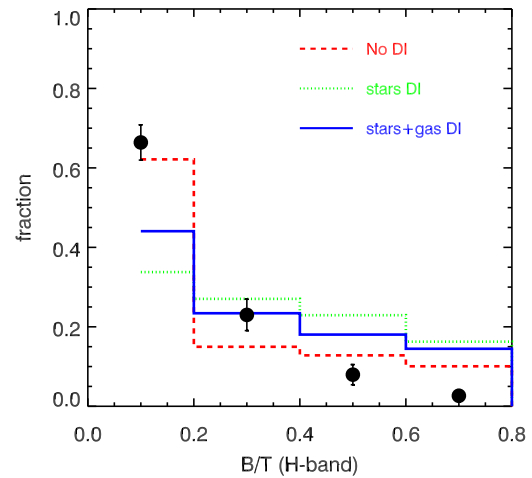}
\caption{The distribution of rest frame $H$-band bulge-to-total ratios
  for disk-dominated galaxies ($B/T<0.75$) with stellar masses greater
  than $10^{10} \msun$. The three colored lines show the `No DI'
  (red), `Stars DI' (green) and `Stars+Gas DI' (blue) models. Symbols
  show the observational estimates of
  \protect\cite{weinzirl:09}. Models with bulge formation via disk
  instabilities underpredict the fraction of disks with very small
  bulges ($B/T<0.2$). }
\label{fig:bttdist}
\end{figure}

However, Figure~\ref{fig:bttdist} shows that our current
implementation of spheroid growth in disk instabilities may not be
completely accurate either. This figure shows the distribution of
bulge-to-total ratios measured in the $H$-band for a sample of
disk-dominated galaxies with $m_* >10^{10} \msun$ by
\citet{weinzirl:09}, compared with our model predictions. As
\citet{weinzirl:09} and other workers have emphasized, about 60\% of
even these relatively massive disks have very small bulges
($B/T<0.2$). Our `No DI' models do reproduce these results, showing
that mergers alone may not produce too many disks with large
spheroids. However, our `stars' and `stars+gas' DI models do not
produce enough disks with $B/T<0.2$ and overproduce objects with
intermediate $B/T$ values.

\begin{figure}
\centering
\includegraphics[width=0.45\textwidth]{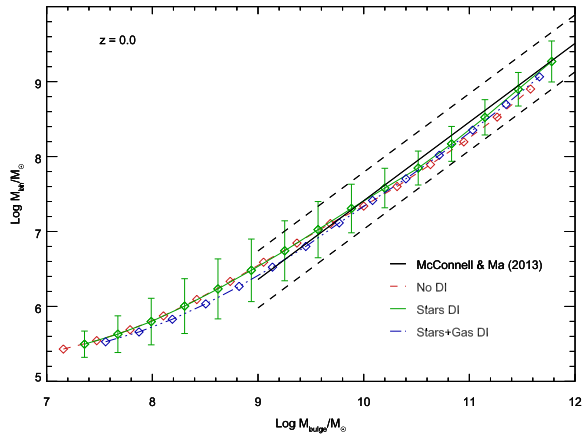}
\caption{The black hole-stellar bulge scaling relation at z = 0.0 for
  the `No DI' (red), `Stars DI' (green) and `Stars+Gas DI' (blue)
  models.  The solid black line shows the $z = 0$ observed relation
  from \protect\cite{McConnell:2013a}, with the dashed lines showing
  the $1-\sigma$ scatter.  The green error bars show the $1-\sigma$
  scatter for the `Stars DI' model and are similar in magnitude to the
  scatter for the other two models.  All three models have been tuned
  to reproduce the normalization of the observed $z = 0$ relation and
  show weak evolution with redshift. The relationship flattens at the
  low mass end due to the $10^{5}\, \msun$ `seed' black holes we have
  adopted in our models.}
		\label{fig:MbhMbulge}

\end{figure}

Figure~\ref{fig:MbhMbulge} shows the black hole-stellar bulge mass
scaling relation for all galaxies at redshift zero.  The three models
produce nearly identical results, falling within the observational
errors of \cite{McConnell:2013a} at redshift zero --- not surprising
as the models have been tuned to match the normalization of the
observed relationship.  The models predict a shallower relation at low
bulge masses.  This is due to the fact that galaxies are `seeded' with
a massive $10^{5}\, \msun $ black hole, creating a floor in the black
hole mass.  We defer a comprehensive study of the evolution of the
black hole scaling relations to future work (Hirschmann et al. in
prep).

\subsection{Growth of the stellar bulge}

\begin{figure*}
\begin{center}
\includegraphics[width=\swidth]{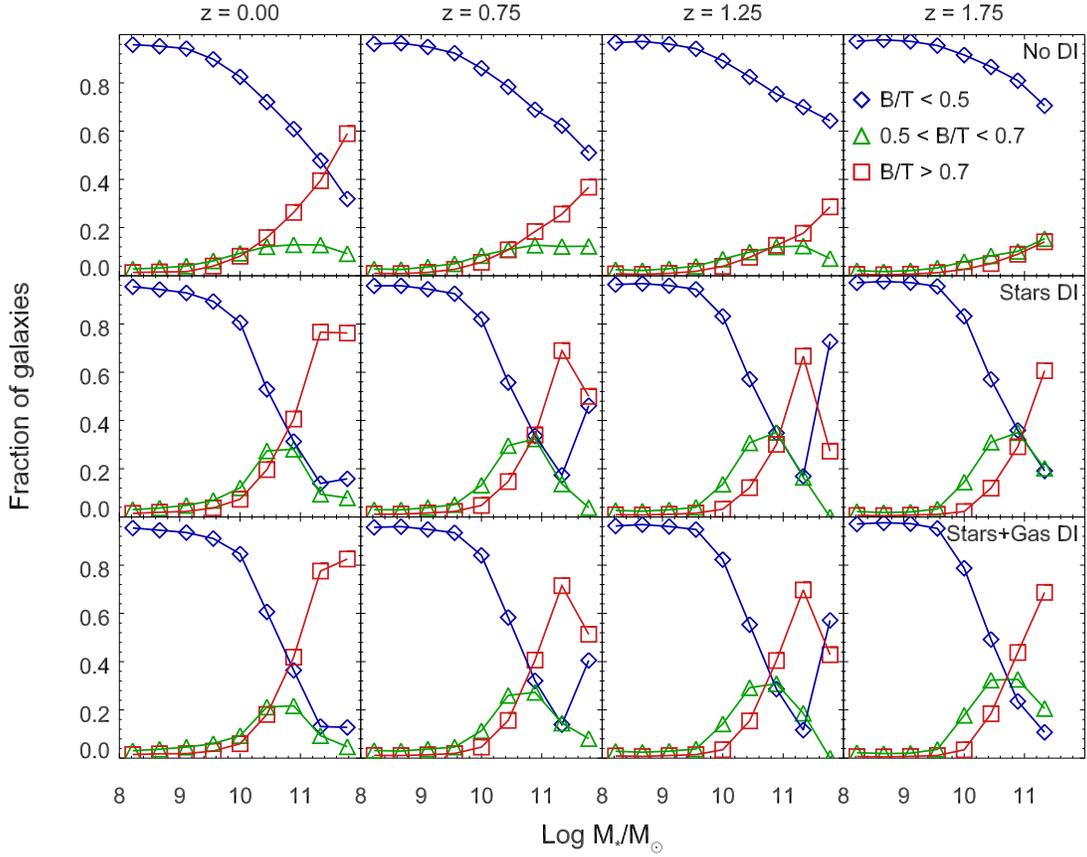}
\caption{Fraction of galaxies with different bulge-to-total ratio as a
  function of stellar mass, at redshifts z = 0.0, 0.75, 1.25, and 1.75
  for the `No DI' (top), `Stars DI' (middle) and `Stars+Gas DI'
  (bottom) versions of the SAM.  The blue, green, and red lines
  indicate the fraction of galaxies of a given mass with stellar
  bulge-to-total ratios between [0,0.5], [0.5,0.7], and [0.7,1.0],
  respectively. All models predict a that a higher fraction of massive
  galaxies are spheroid-dominated; and spheroid-dominated galaxies form
  earlier in the models with disk instabilities.
}
\label{fig:BulgeFraction}
	\end{center}
\end{figure*}

Figure~\ref{fig:BulgeFraction} shows the evolution of the fraction of
galaxies with different values of the bulge-to-total ratio as a
function of stellar mass, from redshift zero to 1.75.  We can see that
not only do the models with disk instabilities produce a larger
fraction of spheroid-dominated galaxies, as we have already seen, but
massive spheroids also form earlier in these models.  In the models
with spheroid growth via disk instabilities, massive galaxies become
spheroid-dominated at high redshifts; above $10^{10.5}\ \msun$ the
majority of galaxies are spheroid-dominated even at z = 1.75, while in
the `No DI' model, only about 30\% of galaxies have $B/T>0.5$.  At low
masses (lower than $10^{10} \msun$) there is no difference between the
two models; disk instabilities evidently have little effect on
low-mass disks in our SAMs.

\subsection{Size-mass relation}
\begin{figure*}
\begin{center}
\includegraphics[width=0.95\textwidth]{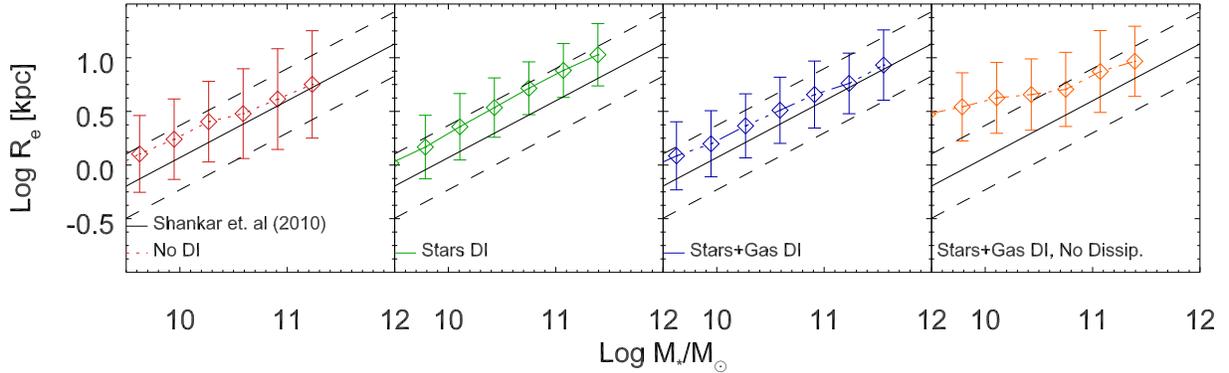}
\label{fig:SizeMassEllip}	
\caption{Size-mass relation for galaxies with $B/T > 0.7$ at redshift
  zero.  The red, green, and blue lines show the `No DI', `Stars DI'
  and `Stars+Gas DI' models with dissipation.  The orange line shows
  the median relation for a model in which all mergers are considered
  to be dissipationless.  The error bars represent the $1\sigma$
  dispersion in the model predictions.  The black line shows the
  observed local relation from \protect\cite{Shankar:2010a}.  All
  dissipational models reproduce the slope of the observed scaling,
  while the dissipationless model produces a relation that is too
  flat, with low mass galaxies having sizes that are too large.  }
\label{fig:sizemassz0}
\end{center}
\end{figure*}

In Figure~\ref{fig:sizemassz0}, we show our model predictions for the
size-mass relation of early-type galaxies at $z\sim 0$.  We compare to
a sample of SDSS galaxies \citep{Hyde:2009a,Shankar:2010a} that were
selected to represent elliptical galaxies and minimize the
contribution by disky S0s; for this reason we limit the analysis to
galaxies with stellar bulge-to-total ratios greater than 0.7.  
All three of our models that include dissipation predict a size-mass
relationship that qualitatively agrees with the slope and dispersion
of the observed relationship, falling within the 1-$\sigma$ error
range for nearly three decades in stellar mass.  We emphasize here
that the model for bulge sizes is never explicitly tuned to
observations; the only free parameters are constrained by
hydrodynamical simulations.  The fact that we predict a local
size-mass relation that is in agreement with observations is thus a
key finding of this paper.

For comparison, we also include a version of the model in which $\Crad
=0.0$ for all mergers.  This dissipationless model produces galaxies
that are too large at all masses, with a size-mass relation that is
nearly flat below $10^{10.5} \msun$.  As described above, the amount
of dissipation is tied to the amount of gas present in the merger.
Since the gas fraction of disk-dominated galaxies increases with
decreasing stellar mass \citep{Kannappan:2004a}, low-mass
spheroid-dominated galaxies are more likely to have formed via gas-rich
processes.  Furthermore, more massive spheroid-dominated galaxies are
more likely to have undergone subsequent dry mergers, weakening the
overall contribution from dissipation.

We now turn to the evolution of the size-mass relation to higher
redshifts.  Numerous observational studies
\citep{Trujillo:2006a,Marchesini:2007a,Toft:2007a, Williams:2010b}
have shown that high-redshift quiescent galaxies are more compact than
their low-redshift counterparts; here we compare to the results of a
recent study by \cite{Newman:2012b}.

\begin{figure*}
\begin{center}
\includegraphics[width=\swidth]{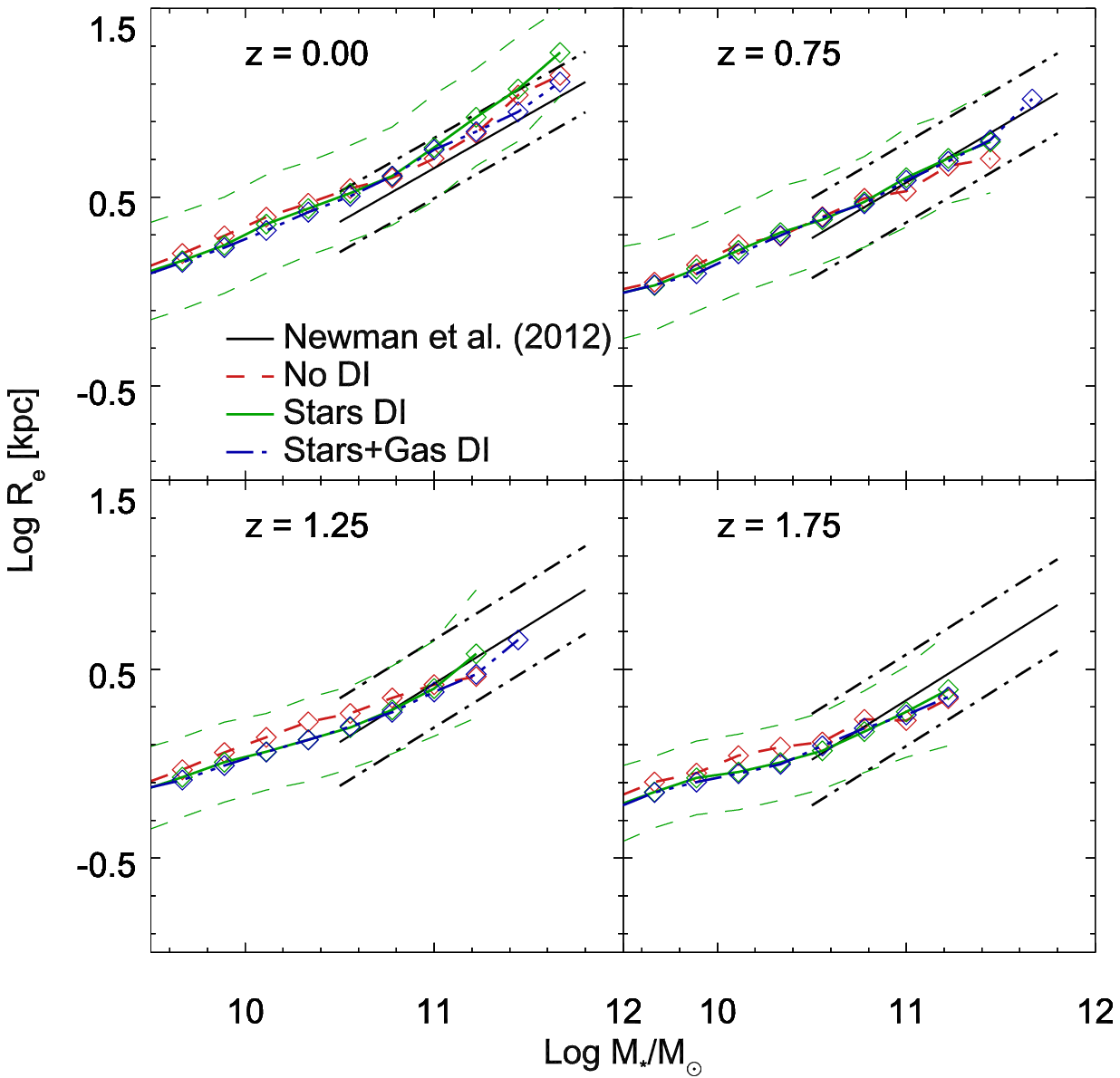}
\caption{Size-mass relation for quiescent galaxies at redshift 0.0,
  0.75, 1.25, and 1.75.  The red, green, and blue lines show the 'No
  DI', `Stars DI' and `Stars+Gas DI' models with dissipation.  The
  green dashed line shows the $1-\sigma$ dispersion in the `Stars DI'
  model; the other two models have similar dispersion.  The black line
  shows the observed relation of \protect\cite{Newman:2012b}.  All
  three models qualitatively reproduce the observed evolution.}
				\label{fig:SizeMassEllipZ}
	\end{center}
\end{figure*}

Following \cite{Newman:2012b}, we select all galaxies with specific
star formation (sSFR) rates less than $0.02\ \rm Gyr^{-1}$.  For
consistency with our other figures we also limit the population to
spheroid-dominated galaxies ($B/T>0.5$), although the results do not
change if we include all quiescent galaxies.  We have converted the
\cite{Newman:2012b} results to a Chabrier IMF for consistency with our
models.  The median size-mass relation at redshifts 0, 0.75, 1.25, and
1.75 is shown in Figure \ref{fig:SizeMassEllipZ}.  

All three models qualitatively reproduce the evolution of the mean
size-mass relation since $z=1.75$.  We obtain similar results if we
compare to \cite{Williams:2010b}, using their evolving sSFR threshold
($\rm sSFR < 0.3\ t_{\rm H}$, where $\rm t_{\rm H}$ is the Hubble time
at that redshift).  We note that similar results are found above
$10^{10.5} \msun$ if we use a mass-weighting instead of a
luminosity-weighting, and if we use the bulge effective radius alone.
Galaxies below $\sim 10^{10.5} \msun $ appear to have a shallower
size-mass relationship above z=1.25. This is below the current limits
of observational samples.  The relation is steeper if we consider the
size of the spheroid alone --- evidently the flattening is due to the
presence of a more extended disk in these low-mass systems. 

\subsection{Faber-Jackson relation and the Fundamental Plane}

Observations have shown that early-type galaxies also fall on a tight
relation between stellar mass and velocity dispersion, termed the
Faber-Jackson \citep[hereafter FJ]{Faber:1976a} relation.  While this
relation is a power law to first order \citep{Gallazzi:2006a}, there
are indications that it may be better approximated by a broken power
law \citep{Tortora:2009a} or a curve
\citep{Hyde:2009a,Cappellari:2012a}, in the sense that more massive
galaxies have relatively lower velocity dispersions.

\begin{figure*}
\begin{center}
\includegraphics[width=\swidth]{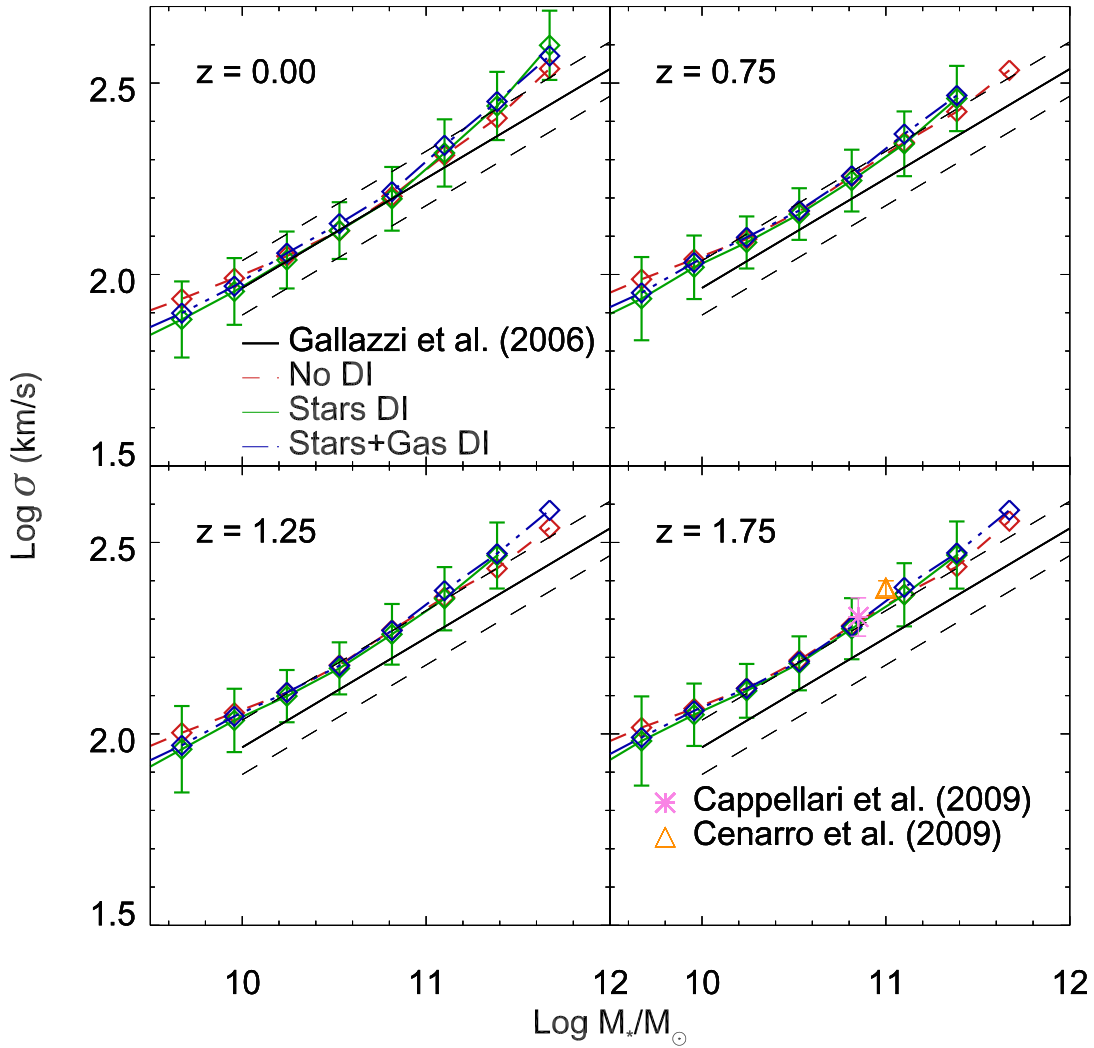}
\caption{Faber-Jackson relation for spheroid-dominated galaxies at
  redshifts z = 0.0, 0.75, 1.25, and 1.75.  The red, green, and blue
  lines show the 'No DI', `Stars DI' and `Stars+Gas DI' models.  The
  green dashed line shows the $1-\sigma$ dispersion in the `Stars DI'
  model; the other two models have similar dispersion.  The black line
  shows the observed $z = 0$ relation, with $1-\sigma$ dispersion
  indicated by the dashed lines \protect\citep{Gallazzi:2006a}.  The
  pink star and orange triangle in the lower right panel show the
  average observed relations at z $\sim$ 1.6
  \protect\citep{Cappellari:2009a,Cenarro:2009a}. While the simulated
  galaxies match the zero-point of the observed low-redshift relation
  they display a curvature in the opposite direction from what is seen
  in observations. This discrepancy could perhaps be understood if the
  IMF varies with galaxy properties such as the velocity dispersion,
  as has been suggested by other studies
  \protect\citep[e.g.][]{Conroy:2012a}.  The simulated galaxies agree
  with observations of the evolution in the normalization of the
  Faber-Jackson relation at higher redshifts.  }
						\label{fig:FaberJackson}
	\end{center}
\end{figure*}

Our predicted FJ relation is shown in Figure \ref{fig:FaberJackson}.
We find that the SAM reproduces the normalization of the relation at
redshift zero \citep{Gallazzi:2006a}.  While the high-redshift FJ
relation is not yet well-constrained by observations, the SAM predicts
that the normalization of the relation increases weakly with redshift, in
agreement with observations \citep{Cappellari:2009a}.

The SAM agrees very well with the predictions of hydrodynamical
simulations \citep{Oser:2012a,Johansson:2012b} and observations of the
evolution of the size-mass and Faber-Jackson relations at fixed
stellar mass
\citep{Trujillo:2006a,Cappellari:2009a,Williams:2010b,Newman:2012b}:
galaxies at higher redshifts have higher velocity dispersions, but
this evolution is much less dramatic than the evolution in the
size-mass relation.  Hydrodynamical simulations
\citep{Dekel:2006a,Robertson:2006b,Hopkins:2010b} have shown that
galaxies that form via gas-rich processes will be compact, with high
velocity dispersions.  Simple analytic arguments \citep{Naab:2009a}
predict that dissipationless minor mergers can greatly increase the
sizes of early-type galaxies while inducing only minor changes in the
velocity dispersion.

In the SAM, this occurs because subsequent minor mergers increase the
effective radius of the galaxy, enclosing more diffuse material within
the effective radius.  Thus the evolution in the velocity dispersion
indicates an evolution in the central surface density of galaxies.
The fact that the SAM reproduces the \emph{magnitude} of the evolution
suggests that the overall SAM framework (i.e. the merger rate, gas
fractions, etc.) can predict the evolution of galaxy properties to
higher redshifts with reasonable accuracy.

While we reproduce the evolution in the normalization of the FJ
relation, our models predict a curvature in the opposite direction to
that seen in local observations.  We predict velocity dispersions that
are too high in the high-mass regime.  This curvature seems to be a
robust feature of our model, and we have checked that our results are
not biased by a population of extremely compact galaxies, or galaxies
too faint to be seen in the local universe.
 
One possibility is that the discrepancy at the high-mass end stems
from variations in the stellar initial mass function (IMF).  There is
a growing body of evidence
\citep{Cappellari:2012a,Conroy:2012a,Dutton:2012b} that early-type
galaxies with high stellar masses or high velocity dispersions may
have a bottom-heavy IMF (e.g. Salpeter), containing more low-mass
stars relative to high-mass stars.  Such an IMF would have a higher
stellar mass-to-light ratio for a stellar population of a given age
and metallicity.

Most observations, such as the \cite{Gallazzi:2006a} analysis, use
stellar luminosities along with stellar population synthesis
techniques to infer the stellar mass.  Using an IMF that is too
top-heavy for high-mass early-type galaxies would tend to underpredict
the stellar masses of galaxies with high velocity dispersions:
\citet{Conroy:2012a} estimate that stellar mass-to-light ratios for a
$\sigma > 300\ \rm{km s^{-1}}$ galaxy may be twice that of the Milky
Way.  Such a correction would raise the high-mass end of the observed
FJ relation by a factor of 0.3 dex. Of course, a non-universal IMF
would also change the results of the SAM, though in a manner that is
difficult to predict. For one thing, the parameters of the SAM would
have to be re-adjusted to fit the revised estimates of the stellar
mass function.

We now turn to the Fundamental Plane, which we consider in the
projection $M_{\rm star} \propto (\sigma^{2} R_{e})^{\alpha}$.  If all
galaxies had the same mass-to-light ratio then one would expect a
simple virial scaling, with $\alpha = 2$.  However observations
\citep{Faber:1987a,Djorgovski:1987b,Dressler:1987a} have shown that
the true FP is tilted from the virial relation, with estimates of the
scaling near $\alpha \sim 1.2$
\citep{Pahre:1998a,Padmanabhan:2004a,Gallazzi:2006a}.  Hydrodynamical
simulations \citep{Dekel:2006a,Robertson:2006a,Covington:2008b} have
shown that if low-mass galaxies have higher gas fractions, they will
experience more dissipation in mergers.  These galaxies will have
higher concentrations of baryonic matter and lower dark matter
fractions within their effective radii, producing a tilt in the
Fundamental Plane. Here we examine this tilt in a cosmological
context.

As there is some evidence that galaxies with large pseudobulges may
have a different FP tilt \citep{Kormendy:2008a} and the observations
we are comparing to either limit their samples to galaxies with high
S\'ersic indices or fit galaxies to a de Vaucouleurs $n=4$ profile, we
restrict this analysis to galaxies that have grown at least half of
their bulge mass through mergers; in practice this restriction has a
minimal effect on the results. For our chosen set of FP variables
there is no dependence on the stellar mass-to-light ratio; any tilt in
the FP comes from variations in the internal structure of the
galaxies.  We find that the SAM closely reproduces the observed local
scaling of $\alpha \approx 1.2$ and, in agreement with observations,
the scatter in the FP is smaller than the scatter in either the
size-mass or Faber-Jackson relations. We find a slight steepening in
the FP at high masses ($M_{*} > 10^{11.5} \msun$), which is generally
consistent with high-redshift observations showing minimal evolution
or a slight increase in the stellar-to-dynamical mass ratio with
redshift \citep{Holden:2010a,Toft:2012a,Bezanson:2013a}.  There is no
dependence on stellar bulge-to-total ratio, and the differences
  between the results of the three model variants ('No DI', 'stars
DI', 'stars+gas' DI) are small.

\begin{figure*}
\begin{center}
\includegraphics[width=\swidth]{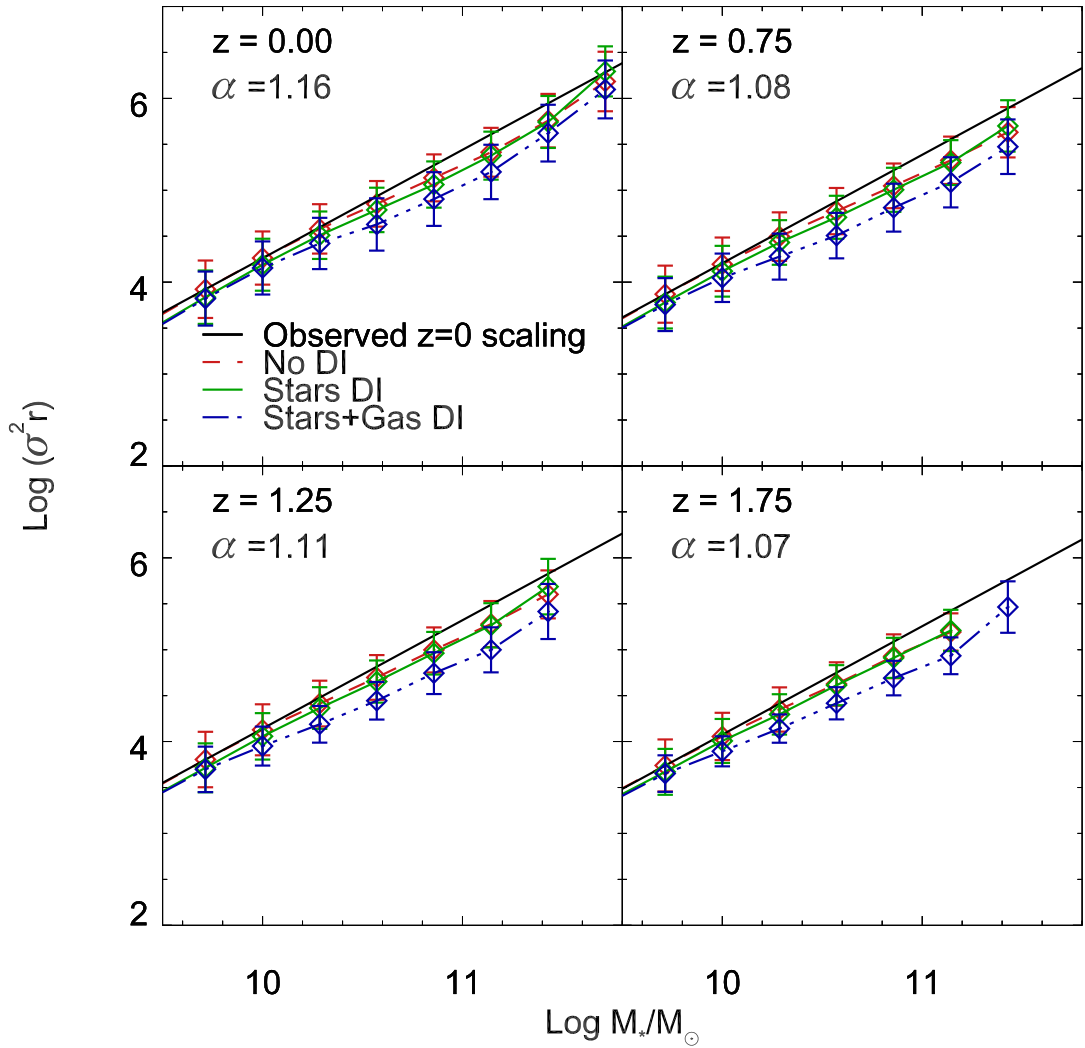}
\caption{Projected Fundamental Plane at redshifts z = 0.0, 0.75, 1.25,
  and 1.75.  The red, green, and blue lines show the 'No DI', `Stars
  DI' and `Stars+Gas DI' models.  Only spheroid-dominated galaxies
  that grew bulges mainly through mergers are shown here. Error bars
  show the 1-$\sigma$ dispersion in the model predictions. The slope
  of the FP in the `Stars DI' model at each redshift is denoted by
  $\alpha$.  In the local universe the SAM reproduces the observed
  tilt of the FP, $\alpha \approx 1.2$, and shows only minor evolution
  towards higher redshift.  Any evolution in the zero-point with
  redshift is within the 1-sigma errors.}
		\label{fig:ProjectedFP}
	\end{center}
\end{figure*}

We find only a minor decrease in the normalization of the FP and a
slightly stronger decrease in the tilt of the FP with redshift.
Performing a least-squares linear fit to the FP relation, we find a
coefficient of $\alpha = 1.21$ at redshift zero, consistent with
observations, evolving to $\alpha = 1.11$ at z\ =\ 1.75.  The
evolution of this tilt supports a scenario in which mergers at lower
redshifts involve less dissipation due to the lower gas content of the
progenitors.  While subsequent dry mergers are thought to preserve the
tilt \citep{Robertson:2006a} induced by the formative event, the
population is continuously supplanted with the products of
increasingly gas-poor mergers.  In this way the tilt of the FP may
slowly evolve over time.

\section{Discussion and Conclusions}

We have updated the ``Santa Cruz'' \citep{Somerville:2008b,
  Somerville:2012a} semi-analytic model to run within accurate merger
trees extracted from the Bolshoi N-body simulation, and to predict the
structural properties of spheroids, using an analytic model
\citep{Covington:2008b,Covington:2011a} based on hydrodynamical
simulations of binary galaxy mergers \citep{Johansson:2009a}.  These
simulations allow us to parameterize the amount of dissipation that
occurs during a merger based on the mass ratio, gas content, size,
morphology, and orbit of the progenitor galaxies.  The hydrodynamical
simulations also show that the velocity dispersion can be determined
simply by using the virial theorem, accounting for the amount of dark
matter in the center of the galaxy. 

In agreement with some previous studies, we find that a model in which
spheroids grow only via mergers has difficulty reproducing the
observed number density of intermediate-mass spheroid-dominated
galaxies at $z=0$. We therefore investigate two different
prescriptions for spheroid growth via disk instabilities, one in which
only stars determine and participate in the disk instability, and one
in which both stars and gas are involved. Massive spheroids form
considerably earlier in both of these models compared with the `No DI'
model, and this may provide a future observational test of the
physical mechanisms included in the models.

All three model variants qualitatively reproduce the slope,
zero-point, scatter and evolution of the size-mass relation for
early-type galaxies. The models with disk instabilities produce better
agreement with these observations than the `No DI' model. We emphasize
that the model has not been tuned to reproduce these observations; the
free parameters in our merger model were instead chosen based on the
results of hydrodynamical binary merger simulations.  We have shown
that the success of this model is due to our treatment of dissipation
in gas-rich mergers; if we neglect dissipation the model predicts a
much flatter size-mass relation, in conflict with observations.
Similar results can be seen in other SAMs that model spheroid sizes
without versus with the inclusion of dissipation
\citep{Guo:2011a,Shankar:2013a}.

\citet{Covington:2011a} presented predictions based on a similar
  model, applied to merger progenitors selected from the Santa Cruz and
  Millennium \citep{Croton:2006a} SAMs in post-processing.  They
  argued that the ``steepening'' of the size-mass relation for early
  types relative to late types, as well as its smaller scatter and
  more rapid evolution, could be qualitatively understood with
  reference to the role of dissipation in mergers, as described in
  their study. If there is a negative correlation between galaxy gas
  fraction and stellar mass, such that lower mass galaxies have higher
  cold gas fractions (as observed), then mergers between lower mass
  galaxies experience more dissipation and result in more compact
  remnants. This will therefore result in a size-mass relation for the
  remnants that is steeper than the progenitor size-mass
  relation. Second, they showed that disk galaxies with larger radius
  at a given stellar mass have higher gas fractions, due to the gas
  surface density dependent star formation efficiency implemented via
  the usual Kennicutt-Schmidt relation. Therefore, larger disks also
  experience more dissipation, leading to a \emph{tightening} in the
  size-mass relation for the remnants relative to the
  progenitors. Third, if galaxy cold gas fractions were higher at high
  redshift, early types formed at lower redshift will experience less
  dissipation, and therefore form more extended remnants, leading to
  evolution in the size-mass relation that is more rapid than that in
  the progenitor population.  They also showed that the observed
  ``tilt'' in the Fundamental Plane can be reproduced in their models;
  the change in tilt relative to the virial relation is due to the
  changing central dark matter fraction that again results from the
  mass dependence of progenitor gas fraction. These are all important
  insights into the possible mechanisms that drive the origin and
  evolution of galaxy structural scaling relations.

However, the Covington et al. study had several important
limitations. It was based on simulations that only considered mergers
between relatively gas rich, disk-dominated progenitors.  Moreover,
the calculations were applied in post-processing on SAM disk-dominated
galaxies that had experienced a recent major merger. Thus all early
type galaxies that formed through mergers at much earlier times were
not included in the populations that were examined. In addition, the
effects of minor mergers, multiple mergers, and mixed-morphology
mergers were neglected. As a result, the Covington approach was not
able to reproduce, for example, the observed stellar mass functions of
disk and spheroid dominated galaxies at the present day. In this work,
we extend the Covington model by calibrating a wider variety of
mergers using a new analysis of the \citet{Johansson:2009a} merger
simulations. We then implement the model \emph{self-consistently}
within the Santa Cruz SAM.

The results of our more detailed and realistic model provide even
stronger support in favor of all of the dissipation-related insights
presented in Covington, summarized above. To these, we add the
important new insight that \emph{major and minor mixed-morphology
  mergers (D-B and B-d) behave like dissipationless mergers.} These
mixed-morphology mergers become increasingly common at redshifts less
than about $z\sim 1.2$, and may play a significant role in the
build-up of the extended envelopes observed in giant elliptical
galaxies in the local universe. The significantly better quantitative
agreement with the evolution of the observed size-mass relation for
early type galaxies that we now obtain with our more sophisticated
model also implies that \emph{several other effects} beyond the ones
already discussed in Covington et al. also contribute to the observed
size evolution, as discussed below. 

It has been suggested that the strong evolution in the size-mass
relation for spheroid-dominated galaxies may be driven largely by
minor mergers \citep[e.g.][]{Naab:2007a,Naab:2009a}. However, some
recent studies have suggested that the minor merger rate may not be
sufficient to account for the observed evolution in the size-mass
relation, particularly above $z \sim 1$
\citep{Newman:2012b,Shankar:2013a}. Our results suggest that several
other effects may contribute to the size evolution as well. The papers
previously mentioned treat only gas-poor mergers as dissipationless.
However, the hydrodynamical simulations we have analyzed show that
mergers between a disk-dominated and a spheroid-dominated galaxy
behave as though they were dissipationless as well.  Since spiral
galaxies tend to be larger than elliptical galaxies at a given mass at
all redshifts, these mergers will greatly increase the effective radii
of early-type galaxies, even more than mergers between two dry
spheroids.  In addition, in our models, high-redshift
spheroid-dominated galaxies can regrow a disk; this process removes
some compact early-type galaxies from the population at lower
redshifts.  The evolution of the gas fraction with redshift also means
that the role of dissipation changes over time, with early, ``wet''
mergers producing more compact remnants, and later, ``dry'' mergers
producing more diffuse spheroids. In a forthcoming paper (Somerville,
Porter et al. in prep), we investigate and elucidate the physical
processes responsible for spheroid growth in our models in more
detail.

Our models predict a slight curvature in the Faber-Jackson relation at
high stellar masses that is not seen in observations, but at lower
stellar masses, the model reproduces the slope and normalization of
the relation.  At higher redshifts we predict a mild increase in
velocity dispersion at fixed mass, in agreement with recent
observations. Our models also match the tilt of the projected
Fundamental Plane ($M_{*} \propto \sigma^{2}r$) at redshift zero.  We
predict a slight decrease in this tilt to higher redshifts; however
the magnitude of this change is well within the measurement errors of
high-redshift observations.

Although we consider the results presented here to be promising, our
existing models have a number of limitations and remaining
uncertainties. Our treatment of disk instabilities is one of the most
uncertain aspects of our models. Both SAMs and hydrodynamical
simulations point to the importance of disk instabilities in forming
spheroids \citep{Dekel:2009a,De-Lucia:2011a,Genel:2012a}, but the
details of the criteria that determine when a disk becomes unstable,
and what happens to it when it does, remain highly
uncertain. Furthermore, our models do not include any environmental
effects such as ram pressure or tidal stripping or harassment, which
may also constribute to galaxy morphological transformation. In
addition, the hydrodynamic simulations upon which we based our recipes
for spheroid formation in mergers are binary mergers of idealized
galaxies that are not in a proper cosmological context. In particular,
they do not have hot gas halos or cosmological accretion. Nor do they
account for ``multiple mergers'' (mergers which occur before a galaxy
has come to equilibrium following a previous merger), which may have
different dynamical effects and are predicted to be common in a \LCDM\
universe \citep{moster:14}. The coming generation of high-resolution
hydrodynamic ``zoom-in'' simulations, including realistic recipes for
stellar feedback, black hole growth, and AGN feedback, confronted with
the results from high-resolution deep imaging surveys, will shed more
light on the physical processes that shape the structure and evolution
of galactic spheroids.

\section{Acknowledgments}
We thank Avishai Dekel, Sandra Faber, Susan Kassin, David Koo, Yu Lu,
Thorsten Naab, and Jerry Sellwood for interesting and useful
conversations.  LP and JP acknowledge funding from NSF-AST 1010033 and
HST-GO-12060.12-A (CANDELS).  PHJ acknowledges the support of the
Research Funds of the University of Helsinki and the Magnus Ehrnrooth
Foundation.

\bibliographystyle{monthly}
\bibliography{Combined.bib}

\end{document}